\documentclass{aa}

\usepackage{graphicx}
\usepackage{txfonts}

\newcommand{\teff}{T_\mathrm{eff}}
\newcommand{\fbol}{f_\mathrm{bol}}
\newcommand{\feh}{\mathrm{[Fe/H]}}
\newcommand{\logg}{\log g}
\newcommand{\afe}{A_\mathrm{Fe}}

\begin{document}

\title{Fundamental Parameters and Abundances of Metal-Poor Stars: The SDSS Standard BD~+17~4708}
\author{I. Ram\'{\i}rez, C. Allende Prieto, S. Redfield\thanks{Hubble Fellow.}, \and D. L. Lambert}
\offprints{I. Ram\'{\i}rez.}
\institute{McDonald Observatory and Department of Astronomy, University of Texas at Austin, RLM 15.306 Austin, TX, 78712-1083 \\ \email{ivan,callende,sredfield,dll@astro.as.utexas.edu}}

\date{Received 21 May 2006 / Accepted 8 Aug 2006}

\abstract{The atmospheric parameters and iron abundance of the \textit{Sloan Digital Sky Survey} (SDSS) spectrophotometric standard star BD~+17~4708 are critically examined using up-to-date Kurucz model atmospheres, LTE line formation calculations, and reliable atomic data. We find $\teff=6141\pm50$~K, $\logg=3.87\pm0.08$, and $\feh=-1.74\pm0.09$. The line-of-sight interstellar reddening, bolometric flux, limb-darkened angular diameter, stellar mass, and the abundances of Mg, Si, and Ca are also obtained: $E(B-V)=0.010\pm0.003$, $\fbol=(4.89\pm0.10)\times10^{-9}$~erg~cm$^{-2}$~s$^{-1}$, $\theta=0.1016\pm0.0023$~mas, $M=0.91\pm0.06M_\odot$, $\mathrm{[Mg/Fe]}=0.40\pm0.10$, $\mathrm{[Si/Fe]}=0.35\pm0.11$, $\mathrm{[Ca/Fe]}=0.36\pm0.11$. This star is a unique example of a moderately metal-poor star for which the effective temperature ($\teff$) can be accurately constrained from the \textit{observed} spectral energy distribution (corrected for reddening). Such analysis leads to a value that is higher than most spectroscopic results previously reported in the literature ($\sim$5950~K). Interstellar reddening was estimated using various prescriptions, including an analysis of interstellar lines. The surface gravity of the star was inferred from the fitting of the wings of the \ion{Mg}{i}\,\textit{b} lines. We used transition probabilities measured in the laboratory and reliable damping constants for unblended Fe lines to derive the iron abundance using both \ion{Fe}{i} and \ion{Fe}{ii} lines. We find that the ionization balance of Fe lines is satisfied only if a low $\teff$ ($\sim5950$~K) is adopted. The mean iron abundance we obtain from the \ion{Fe}{ii} lines corresponds to $A_\mathrm{Fe}=5.77\pm0.09$ ($\feh=-1.74$ for our derived $A_{\mathrm{Fe},\odot}=7.51$) while that from the \ion{Fe}{i} lines is $A_\mathrm{Fe}=5.92\pm0.11$, and therefore with our preferred $\teff$ ($6141$~K), the discrepancy between the mean iron abundance from \ion{Fe}{i} and \ion{Fe}{ii} lines cannot be explained by overionization by UV photons as the main non-LTE effect. Interestingly, the \ion{Fe}{i} excitation balance is satisfied with a $\teff$ only slightly warmer than our preferred solution and not with the lower value of 5950~K. We also comment on non-LTE effects and the importance of inelastic collisions with neutral H atoms in the determination of oxygen abundances in metal-poor stars from the 7774~\AA\ \ion{O}{i} triplet.
}

\titlerunning{The SDSS standard BD~+17~4708}
\authorrunning{Ram\'{\i}rez et~al.}
\maketitle

\section{Introduction}

The derivation of stellar chemical compositions relies on the accurate determination of the atmospheric parameters $\teff$ (effective temperature) and $\logg$ (surface gravity). These quantities may be inferred either from the stellar spectrum or by semi-empirical methods that are normally based on photometric and astrometric measurements. Often, the photometric and astrometric parameters are used as first approximations and the final solution is found iteratively with the help of the spectrum. This tuning of parameters is, however, model-dependent, and may lead to erroneous conclusions if the models are inadequate.

Most abundance analyses of FGK stars are made using homogeneous plane-parallel model atmospheres and LTE (local thermodynamic equilibrium) line formation. However, recent abundance analyses using hydrodynamical model atmospheres and non-LTE line formation have demonstrated that the effects of surface inhomogeneities and departures from LTE on abundance analyses are not negligible in the Sun and solar-type stars of different metallicities (e.g., Asplund \& Garc\'{\i}a~Perez 2001, Korn et~al. 2003, Allende~Prieto et~al. 2004b, Asplund 2005), being probably dramatic for very metal-poor stars (Shchukina et~al. 2005).

The F8-type star BD~+17~4708 has been chosen as a spectrophotometric standard, either primary or secondary, for various systems (e.g., Oke \& Gunn 1983, Rufener \& Nicolet 1988, J\o rgensen 1994, Zhou et~al. 2001). In particular, this star is the primary standard of the \textit{Sloan Digital Sky Survey} (SDSS) photometric system (Fukugita et~al. 1996; Gunn et~al. 1998, 2006; Smith et~al. 2002). BD~+17~4708 is one of the few stars, and the only subdwarf, with very accurate absolute fluxes (Bohlin \& Gilliland 2004b), which allows us to test, in an independent way, the models and different atmospheric parameters that have been derived for it. In particular, it is a unique case in which the effective temperature may be well constrained. Also, it can be quite useful to know, with high accuracy, what the fundamental parameters and overall chemical composition of this star are, given that its model atmosphere and predicted fluxes may be used to complement the observed spectral energy distribution in the transformation of observed magnitudes into physical fluxes in the SDSS.

\begin{figure}
 \centering
 \includegraphics[width=7.0cm]{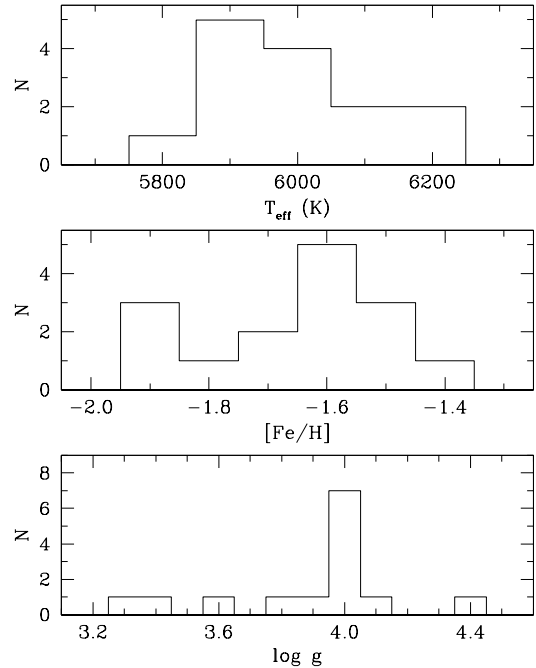}
 \caption{Distribution of the $\teff$, $\feh$, and $\logg$ values found in the literature for the star BD~+17~4708, as given in Table~\ref{t:literature}.}
 \label{f:literature}
\end{figure}

BD~+17~4708 has been studied by several groups (Table~\ref{t:literature}), who have derived effective temperatures between 5800~K and 6200~K, $\feh$ values between $-2.0$ and $-1.4$ and a surface gravity of $\logg\simeq4.0$ (Fig.~\ref{f:literature}).\footnote{Throughout this paper, we use the standard definitions $\mathrm[X/Y]=\log(N_X/N_Y)-\log(N_X/N_Y)_\odot$ and $A_\mathrm{X}=\log(N_X/N_H)+12$, where $N_X$ is the number density of the element $X$. The surface gravity $g$ in $\logg$ is given in cgs units.} In this paper, we present a detailed determination of the atmospheric parameters and the iron abundance of BD~+17~4708. A critical comparison with values previously reported in the literature is provided in the Appendix. Due to their relevance for studies of stellar interiors, we also derive the abundances of Mg, Si, and Ca, and discuss the uncertainties in the determination of the oxygen abundance from the 7774~\AA\ triplet, in particular the importance of inelastic collisions with neutral H atoms in the non-LTE computations involved. Our analysis is based on the observed spectral energy distribution and a high resolution, high signal-to-noise (S/N) spectrum of the star.

\begin{table}
 \centering
 \begin{tabular}{lccc}
  \hline\hline
   Reference & $\teff$ (K) & $\logg$ & $\feh$ \\ \hline
   Peterson (1981)                & 5800 & 4.00 & $-1.95$ \\
   Rebolo et~al. (1988)           & 5890 & 4.00 & $-1.70$ \\  
   Magain (1989)                  & 5960 & 3.40 & $-1.93$ \\
   Axer et~al. (1994)             & 6100 & 4.40 & $-1.42$ \\
   Spite et~al. (1994)            & 5950 & 3.30 & $-1.50$ \\
   Th\'evenin \& Idiart (1999)    & 5929 & 4.02 & $-1.54$ \\
   Boesgaard et~al. (1999, K93)   & 6091 & 3.81 & $-1.73$ \\
   Boesgaard et~al. (1999, C83)   & 5956 & 3.65 & $-1.81$ \\
   Fulbright (2000)               & 6025 & 4.00 & $-1.63$ \\
   Mishenina et~al. (2000)        & 6000 & 4.00 & $-1.65$ \\
   Ryan et~al. (2001)             & 5983 & ---- & $-1.86$ \\
   Simmerer et~al. (2004)         & 5941 & 3.98 & $-1.60$ \\
   Nissen et~al. (2004)           & 5943 & 3.97 & $-1.64$ \\
   Mel\'endez \& Ram\'{\i}rez (2004) & 6154 & 3.93 & $-1.64$ \\   
   Asplund et~al. (2006)          & 6183 & 4.11 & $-1.51$ \\   
   \hline
 \end{tabular}
 \caption{Literature data for BD~+17~4708. Note that there are two sets of parameters given by Boesgaard et~al. (1999): those given in the King (1993, K93) scale, and those given in the Carney (1983a,b; C83) scale. In this Table, only values derived for the first time (although mostly taken from the more recent papers) by each author(s) are shown to avoid duplicity.}
 \label{t:literature}
\end{table}

\section{Fitting of the spectral energy distribution} \label{s:fluxfit}

Bohlin \& Gilliland (2004b, hereafter BL04b) have measured the spectral energy distribution of BD~+17~4708, from the UV (0.17~$\mu$m) to the near IR (1.0~$\mu$m), with respect to the three primary standards of the \textit{Space Telescope Imaging Spectrograph} (STIS) on the \textit{Hubble Space Telescope} (HST). The accuracy of BL04b measurements is better than 0.5\% and therefore, in the absolute scale, their fluxes for BD~+17~4708 are as accurate as those of the three white dwarfs used as primary standards. The fluxes measured with STIS for these three white dwarfs have uncertainties that range from 2.5\% in the UV to 1\% at longer wavelengths, according to BL04b. Thus, the spectral energy distribution of BD~+17~4708 is accurate at the level of 2\% or even better in some spectral regions.

\begin{figure*}
 \centering
 \includegraphics[width=13cm]{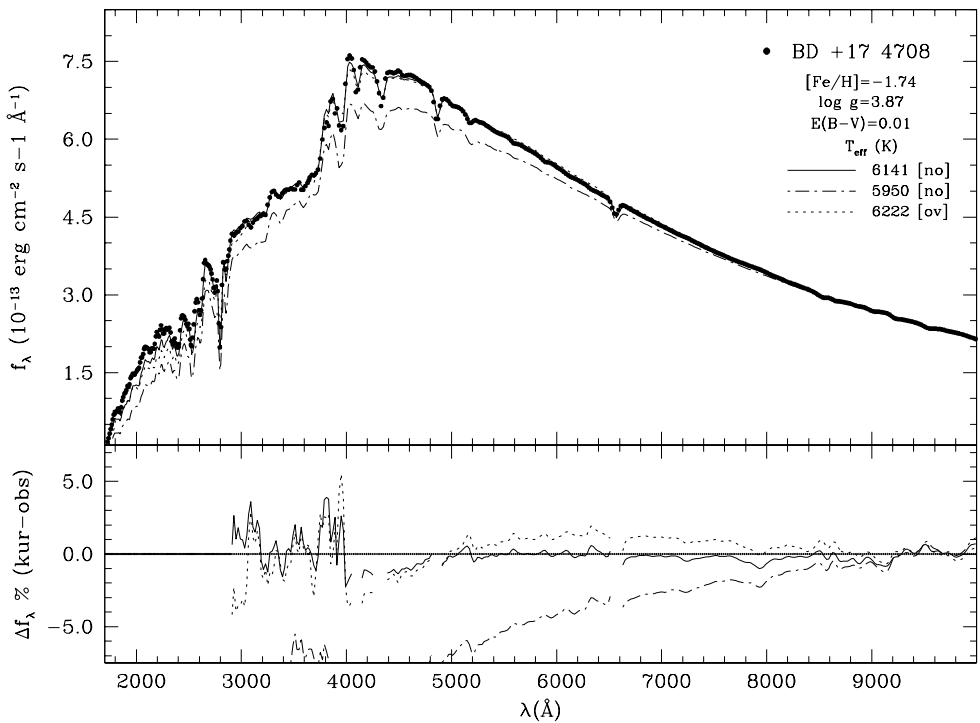}
 \caption{Upper panel: The spectral energy distribution measured by Bohlin \& Gilliland (2004b) is shown with the filled circles. Alpha-element enhanced ([$\alpha$/Fe]=+0.4) no-overshoot [no] Kurucz's models of $\feh=-1.74$, $\logg=3.87$, reddened by $E(B-V)=0.01$ according to the Fitzpatrick (1999) parameterization, and two effective temperatures: $\teff=6141$~K and 5950~K, are shown with the solid and dot-dashed lines, respectively. An overshoot [ov] Kurucz model of $\teff=6222$~K is shown with the dotted line ($\logg$ and $\feh$ for this model are slightly different to make them consistent with the higher $\teff$). Lower panel: As in the upper panel for the difference (percentile) between observed and theoretical fluxes. The hydrogen lines have been excluded.}
 \label{f:flux}
\end{figure*}

In Fig.~\ref{f:flux}, the observed fluxes from BL04b are shown along with theoretical flux distributions computed by R.~L.~Kurucz\footnote{Up-to-date models are available at http://kurucz.harvard.edu. The characteristics of the models are explained in Kurucz (1970, 1979).} after applying an $E(B-V)=0.01$ reddening (see \S\ref{s:reddening}) according to the Fitzpatrick (1999) parameterization with $R_V=A_V/E(B-V)=3.1$. The models have $\logg=3.87$ and $\feh=-1.74$, values that we derive in~\S\ref{s:lines}. The theoretical fluxes have been empirically scaled to the observed one using the reddest 1000~\AA, i.e., they were divided by the mean ratio of theoretical to observed fluxes from 9000~\AA\ to 10000~\AA\ in each case. The scaling factor, $s$, is directly related to the stellar angular diameter, $\theta$, by $s=\theta^2/4$. The spectrum from BL04b was smoothed to approximately match the resolving power of the Kurucz model fluxes.

As shown in Fig.~\ref{f:flux}, up-to-date Kurucz models, which include newly computed opacity distribution functions, an $\alpha$-element enhancement consistent with the mean [$\alpha$/Fe] ratio we obtain (\S\ref{s:alpha}), and without convective overshooting, accurately reproduce (within 1\%) the observed flux distribution at wavelengths longer than 5000~\AA, as long as the correct $\teff=6141$~K is adopted (\S\ref{s:teff}). In the range 4000~\AA$<\lambda<5000$~\AA\ the model underestimates the flux by about 1.5\% while at shorter wavelengths the fit is reasonable on average although several strong lines are not well fitted. Interestingly, a comparison of a MARCS model\footnote{Downloaded from http://marcs.astro.uu.se.} with a Kurucz model of parameters near those of BD~+17~4708 (the closest point in the grid we found is: $\teff=6000$~K, $\logg=4.0$, $\feh=-2.0$) shows that the MARCS model predicts fluxes in this region that are larger by about 1\%, which would reduce the difference somewhat. Kurucz and MARCS models seem to predict roughly the same fluxes everywhere else.

The Kurucz overshoot model shown in Fig.~\ref{f:flux} has the $\teff$ that best fits the data for overshoot models. Clearly, it does not reproduce very well the observational data. The overshoot model does not have $\alpha$-element enhancement but this has a much smaller effect on the shape of the spectral energy distribution compared to the switch to a no-overshoot model. Note also that the overshoot model that fits best the observations (that shown in Fig.~\ref{f:flux}) is hotter by about 80~K compared to the no-overshoot model, thus introducing a systematic error in the $\teff$ derived from these fits. For that reason, we prefer to adopt Kurucz no-overshoot models hereafter. Note also that adopting a Kurucz model of $\teff=5950$~K, as suggested by previous spectroscopic studies (Table~\ref{t:literature} and Fig.~\ref{f:literature}), results in a severe discrepancy with the observational data.

Although it is not shown in Fig.~\ref{f:flux}, there is a slight degeneracy between $\teff$ and $E(B-V)$ in the model fits to the observed flux distribution. Roughly speaking, increasing $\teff$ is equivalent to decreasing $E(B-V)$, as they both result in higher fluxes in the UV-blue regions while leaving the red and infrared fluxes nearly unchanged (after scaling). For instance, equally good fits to the data can be obtained with $\teff\simeq6050$~K if no reddening is assumed or with $\teff\simeq6150$~K if $E(B-V)=0.010$ is adopted. It is, thus, important to constrain the $E(B-V)$ value independently.

\subsection{Reddening} \label{s:reddening}

Since BD~+17~4708 is at a distance of about 120~pc (its \textit{Hipparcos} parallax is $8.43\pm1.42$~mas), the $E(B-V)$ value is expected to be negligible or small.  The Local Bubble, a region devoid of dense gas extends approximately 60~pc in the direction of BD~+17~4708 (Lallement et~al. 2003).  At the Local Bubble boundary, a significant increase in dense interstellar medium (ISM) material is observed toward several stars in the general direction of BD~+17~4708. This ISM material can be observed as narrow absorption lines in high resolution spectra, and depending on the strength of the absorption, it may be expected to cause a small, but measurable, amount of reddening.

\begin{figure*}
 \centering
 \includegraphics[width=13.5cm]{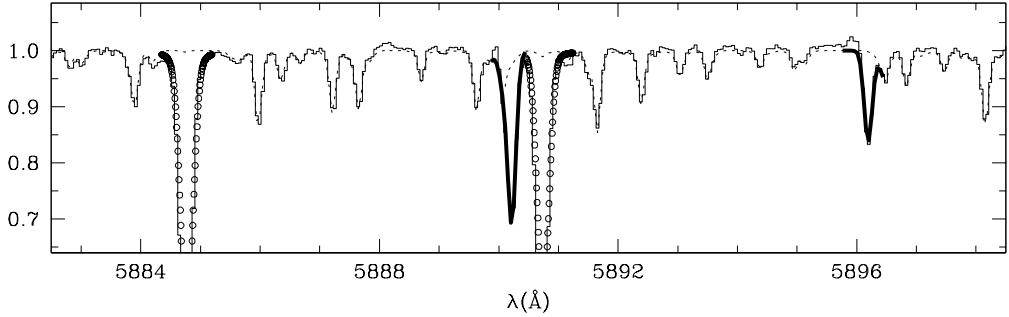}
 \caption{The observed spectral region around the Na D lines for BD~+17~4708, uncorrected for the radial velocity of the star, is shown with the histogram (these observational data are described in \S\ref{s:observations}). The dotted line is the model for telluric water vapor, open circles correspond to the model fit to the stellar lines (see \S\ref{s:gravity}), and the thick solid line is the model fit to the interstellar \ion{Na}{D} lines (including blends with telluric lines).}
 \label{f:ism}
\end{figure*}

Interstellar gas in the line of sight of BD~+17~4708 is evidenced by the interstellar \ion{Na}{i} D1 ($\sim5896$~\AA) and D2 ($\sim5890$~\AA) lines shown in Fig.~\ref{f:ism}.  Due to the high radial velocity of the star ($-291$~km~s$^{-1}$), the ISM absorption lines at $-13$~km~s$^{-1}$ are significantly displaced from the strong stellar features.  Other than the two stellar lines and two interstellar lines, all of the remaining features seen in Fig.~\ref{f:ism} are caused by telluric water vapor.

Although the telluric H$_2$O lines are relatively weak, they need to be modeled and removed from the spectrum, in order to obtain a high precision measurement of the \ion{Na}{i} ISM column density. A relatively simple model of terrestrial atmospheric transmission (AT - Atmospheric Transmission program, from Airhead Software, Boulder, CO) developed by Erich Grossman is used to fit the telluric water vapor lines. This forward modeling technique to remove telluric line contamination in the vicinity of the \ion{Na}{i} D lines is described in detail by Lallement et~al. (1993), in which a more sophisticated terrestrial atmospheric model was employed. As can be seen in Fig.~\ref{f:ism}, the AT program is very successful at modeling the terrestrial absorption in the spectrum of BD~+17~4708.  Note that the stellar lines happen to fall in an area free of contaminating lines, and the two interstellar lines are only slightly blended with water vapor absorption in the wings of the ISM absorption.

The interstellar lines found in the BD~+17~4708 spectrum were modeled using standard methods (see, e.g., \S2.2 in Redfield \& Linsky 2004a). A single Gaussian absorption component is fit to both \ion{Na}{i} D lines simultaneously using atomic data from Morton (2003), and then convolved with the instrumental line spread function. Fitting the lines simultaneously reduces the influence of systematic errors, such as continuum placement and contamination by weak features. The free parameters are the central velocity ($v$), the line width or Doppler parameter ($b$), and most importantly, because it can be used to estimate the reddening along the line of sight, the column density ($N$) of \ion{Na}{i} ions toward BD~+17~4708. The best fit is shown in Fig.~\ref{f:ism}, where $v=-13.315\pm0.031$~km~s$^{-1}$, $b=3.922\pm0.052$~km~s$^{-1}$, and $\log N(\ion{Na}{i})=11.4776\pm0.0034$~cm$^{-2}$.  Due to the high S/N, systematic errors probably dominate over the statistical errors given above.  

The measured line width, a consequence of thermal and non-thermal, or microturbulent, broadening (see Redfield \& Linsky 2004b), is wider than typically found for cold ISM clouds (Welty et~al. 1994). This is likely due to unresolved interstellar components along the line of sight. Although this observation is considered to be high spectral resolution from a stellar perspective ($\lambda/\delta\lambda\simeq60,000$, see \S\ref{s:observations}), from an interstellar perspective it is moderate resolution because the narrow and closely spaced component structure typical of the ISM is best observed at much higher resolution ($\lambda/\delta\lambda\simeq500,000-1,000,000$). Therefore, we are likely seeing blending of several ISM components along the line of sight toward BD~+17~4708. In fact, $\pi$\,Aqr, a star in the same part of the sky as BD~+17~4708 ($\Delta\theta\sim15^{\circ}$), but more distant, has been observed at high resolution ($\lambda/\delta\lambda\simeq600,000$) by Welty et~al. (1994).  They observed, among 8 total interstellar components, three with velocities between $-11$ and $-13.7$~km~s$^{-1}$, with a column density weighted average velocity of $-12.5$~km~s$^{-1}$.  The total column density for these three components is $\log N(\ion{Na}{i})=12.20$~cm$^{-2}$. Because $\pi$\,Aqr is more distant than BD~+17~4708, several more components are observed, and a higher column density is to be expected, but the agreement between the two lines of sight, further confirms the interstellar origin of the absorption lines observed in the BD~+17~4708 spectrum.  

Although we are likely observing several blended ISM components toward BD~+17~4708, because the absorption is optically thin, a single component fit to the entire absorption feature should provide a precise measurement of the total \ion{Na}{i} column density. It has been shown that, even though most of the ISM Na is ionized, the total \ion{Na}{i} column density correlates very well with the total hydrogen column density, $N(\ion{H}{i}+H_2)$ (Ferlet et~al. 1985). Using the relation between $N(\ion{Na}{i})$ and $N(\ion{H}{i}+H_2)$ provided by Ferlet et~al. (1985), which holds for \ion{Na}{i} column densities in the range $10.0\leq \log N(\ion{Na}{i})$~cm$^{-2}\leq13.0$, we derive a total hydrogen column density of $\log N(\ion{H}{i}+H_2)=19.78$~cm$^{-2}$. Bohlin et~al. (1978) provide a calibration to transform a total hydrogen column density into an $E(B-V)$ value, which for the highest extinctions has been confirmed by Rachford et~al. (2002). Using this relation we obtain a reddening value of $E(B-V)=0.010$.  

No formal error bars are provided for the transformations from $N(\ion{Na}{i})$ to $N(\ion{H}{i}+H_2)$ or $N(\ion{H}{i}+H_2)$ to $E(B-V)$. However, we expect the estimate of total hydrogen column density from $N(\ion{Na}{i})$ to be very good because our observed \ion{Na}{i} column density is right in the middle of the distribution of points Ferlet et~al. (1985) used to calibrate this relation. The transformation from $N(\ion{H}{i}+H_2)$ to $E(B-V)$ is more difficult because we are at the lower end of the distribution of measurements used by Bohlin et~al. (1978). At these low reddenings, there is some dispersion in the relation due to the low number of discrete absorbers along short sightlines, whereas the relation is significantly tighter for more distant lines of sight, over which a much larger number of ISM environments are averaged. The lowest hydrogen column densities used by Bohlin et~al. (1978) are comparable to the column density we observe toward BD~+17~4708. Those targets with similar hydrogen column densities have measured reddenings in the range of $E(B-V)=0.01$ to 0.02.

\bigskip

Other methods to estimate $E(B-V)$ were also used. The $E(B-V)$ value from the Schuster \& Nissen (1989) calibration, which is based on Str\"omgren photometry, is essentially zero. Interstellar extinction surveys by Fitzgerald (1968) and Arenou et~al. (1992) suggest $E(B-V)=0.000$ and 0.017, respectively, when used in conjunction with the Hakkila et~al. (1997) code, which takes into account the distance to the star. The empirical laws by Bond (1980) and Chen et~al. (1998) suggest $E(B-V)=0.024$ while the integrated extinction maps by Burstein \& Heiles (1978) and Schlegel et~al. (1998) set upper limits of $E(B-V)=0.043$ and 0.035, respectively. Note, however, that all the map estimates have large error bars. A simple mean of the $E(B-V)$ from the maps results in $0.014\pm0.010$.

\begin{figure}
 \centering
 \includegraphics[width=7.5cm]{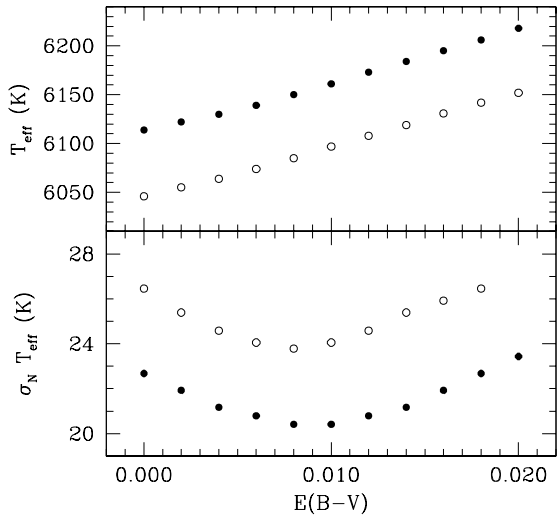}
 \caption{Top panel: mean $\teff$ obtained from the $\teff$-color calibrations by Ram\'{\i}rez \& Mel\'endez (2005b) as a function of $E(B-V)$ using all colors (14, open circles), and blue-visible colors only (7, filled circles). Bottom panel: as in the top panel for the standard error $\sigma_N=\sigma\teff/\sqrt{N}$, where $\sigma$ is the standard deviation and $N$ the number of colors used.}
 \label{f:teffcolor}
\end{figure}

Another way to estimate $E(B-V)$ is by means of the use of several homogeneously calibrated unreddened color-temperature relations. In principle, the standard deviation from the mean of several color temperatures minimizes when the appropriate $E(B-V)$ value is used. Thus, we used 14 of the color-temperature relations by Ram\'{\i}rez \& Mel\'endez (2005b) and found the dispersion from the mean $\teff$ to be a minimum with $E(B-V)=0.008\pm0.001$ (see Fig~\ref{f:teffcolor}). The real error bar in this $E(B-V)$ value is very likely to be larger due to photometric uncertainties and systematic errors in the $\teff$-color calibrations. Also, the presence of a cool companion (see \S\ref{s:binarity}) may be affecting this estimate by increasing the red and infrared fluxes compared to the case of a single star. In fact, using only blue-visible colors we obtain $E(B-V)=0.009\pm0.002$ (Fig.~\ref{f:teffcolor}).

\bigskip

In summary, BD~+17~4708 is slightly affected by interstellar reddening but the exact $E(B-V)$ is uncertain. Interstellar extinction maps suggest $E(B-V)=0.014\pm0.010$. The $E(B-V)$ value from the $\teff$-color relations has a more reasonable error of about 0.003, including systematic errors. If we assume a slightly smaller error to the $E(B-V)$ value from the fit to the ISM lines, e.g., 0.002, we obtain a weighted mean of $E(B-V)=0.010\pm0.003$. The independent error bars given here are somewhat arbitrary but appropriate for the estimate of the weighted mean.

\subsection{Effective temperature} \label{s:teff}

The BL04b data provides a very reliable way to determine the effective temperature of BD~+17~4708 with the help of theoretical flux distributions, provided $\logg$, $\feh$, and $E(B-V)$ are known with sufficient accuracy. In particular, the high sensitivity of the UV continuum flux to $\teff$ can be used to constrain $\teff$ to a level of about 100~K. Even though the completeness of the UV continuum and line opacities in the Kurucz models is controversial (Bell et~al. 1994, Balachandran \& Bell 1998), the observed UV fluxes have been shown to be reasonably well reproduced by Kurucz models in the Sun (Allende~Prieto et~al. 2003b), Vega (Bohlin \& Gilliland 2004a, Garc\'{\i}a-Gil et~al. 2005), and late-type stars of different metallicities (Allende~Prieto \& Lambert 2000). We did not include the far UV ($\lambda<2900$ \AA) to quantify the quality of the fits in Fig.~\ref{f:flux}. Not only does this avoid possible errors in the UV fluxes of Kurucz models but it also reduces the impact of errors in the observed flux distribution, which is less accurate at short wavelengths. Nonetheless, including the far UV does not change our conclusions significantly. We also excluded the strong hydrogen lines because they are affected by non-LTE.

\begin{figure}
 \centering
 \includegraphics[width=7.5cm]{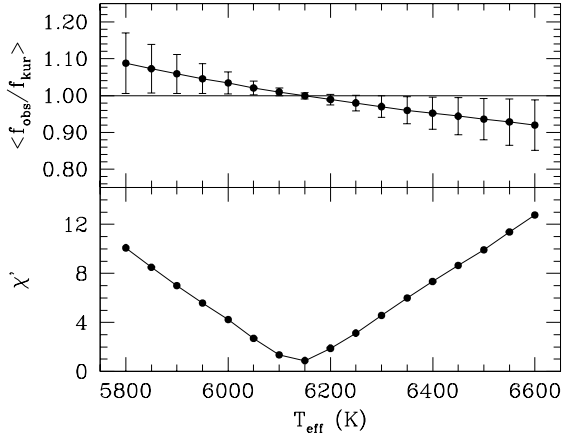}
 \caption{Upper panel: mean value of the ratio of observed ($f_\mathrm{obs}$) to scaled theoretical ($f_\mathrm{kur}$) fluxes (see Fig.~\ref{f:flux}) as a function of $\teff$. Bottom panel: the quality of the fit, as given by $\chi'=\sqrt{\chi^2/(n-1)}$ where $\chi^2=\sum_{i=1}^n[(f_\mathrm{obs}-f_\mathrm{kur})_i^2/\sigma_i^2]$, as a function of $\teff$ ($\sigma$ is the error in $f_\mathrm{obs}$ only).}
 \label{f:fancy}
\end{figure}

As shown in Fig.~\ref{f:flux}, the UV and blue-visible spectral regions are not well fitted with the $\teff=5950$~K model. Any attempt to reconcile the model and observed UV continuum fluxes with an increase in the scaling factor ruins the agreement in the infrared. The best fit to the data is obtained with $\teff=6141$~K for $E(B-V)=0.01$ (see Fig.~\ref{f:fancy}). Given the error in the extinction value derived in \S\ref{s:reddening}, the temperature of BD~+17~4708 is well constrained, from the flux fit, at the 50~K level, i.e., $\teff=6141\pm50$~K. This 50~K error includes the uncertainties in the other atmospheric parameters ($\logg$, $\feh$) but it is still dominated by the error in the $E(B-V)$ value. A systematic error due to the choice of models is certainly present but not included in the 50~K.

Most previous spectroscopic studies of BD~+17~4708 have found a low $\teff$ of about 5950~K (see Table~\ref{t:literature} and Fig.~\ref{f:literature}).\footnote{By `spectroscopic' here we refer to $\teff$ obtained from the excitation and/or ionization balance of iron lines.} In the Cayrel~de~Strobel et~al. (2001) catalog, for example, the 8 entries found for this star have $\teff$ between 5790~K and 6100~K, with the mean $\teff$ being 5960~K. Application of the InfraRed Flux Method (IRFM) for BD~+17~4708 resulted in 5955~K according to Ram\'{\i}rez \& Mel\'endez (2005a), who adopted $E(B-V)=0.000$. Note, however, that if $E(B-V)=0.010$ is adopted, the IRFM color-temperature calibrations by Ram\'{\i}rez \& Mel\'endez (2005b) suggest $\teff\simeq6100$~K and if we use blue-visible colors only we obtain $\teff\simeq6150$~K (Fig.~\ref{f:teffcolor}). The low temperature obtained from the IRFM and red/infrared photometry is probably due to the presence of a cool companion (see \S\ref{s:binarity}). The three highest temperatures in Table~\ref{t:literature} are those by Axer et~al. (1994), Mel\'endez \& Ram\'{\i}rez (2004), and Asplund et~al. (2006). Axer et~al. and Asplund et~al. derived their $\teff$ from fitting the wings of the Balmer lines, while Mel\'endez \& Ram\'{\i}rez used several IRFM temperature-color relations with a relatively high $E(B-V)\simeq0.02$ value.

\subsection{Bolometric flux}

The observed absolute flux curve by BL04b covers spectral regions that are difficult to model, namely the UV and blue-visible. Beyond 1.0~$\mu$m, the flux distribution is very well behaved and accurately reproduced by the stellar atmosphere models. Furthermore, the infrared portion of the spectrum is almost insensitive to the choice of effective temperature when the theoretical fluxes are normalized at the Rayleigh-Jeans tail.

The bolometric flux of BD~+17~4708 was obtained by integrating the observed flux distribution up to 1.0~$\mu$m and the predictions from the models for longer wavelengths. Note that for the model fits, as in Fig.~\ref{f:flux}, the theoretical spectra are the ones that have been reddened. For the bolometric flux calculation, on the other hand, we unreddened the observed flux distribution. Thus, we use the term bolometric flux in the intrinsic sense, i.e., we use it to refer to the flux that would be measured at the top of the Earth's atmosphere in the absence of interstellar absorption.

The mean error of each point on the observed flux distribution is about 2\%. Assuming the error in the models is negligible, our best estimate for the bolometric flux is $\fbol=4.89\pm0.10\times10^{-9}$~erg~cm$^{-2}$~s$^{-1}$. This value is in good agreement with that obtained from the Alonso et~al. (1995) photometric calibrations for the $(K,V-K)$ pair, which result in $4.80\times10^{-9}$~erg~cm$^{-2}$~s$^{-1}$.

\subsection{Angular diameter} \label{s:theta}

The angular diameter of BD~+17~4708 can be calculated with the $f_\mathrm{bol}$ and $\teff$ derived above. This value corresponds to the limb-darkened angular diameter. Propagating the 2\% error in $f_\mathrm{bol}$ and 1\% error in $\teff$ we determine our best solution for the angular diameter as: $\theta=0.1016\pm0.0023$~mas.

\subsection{Binarity} \label{s:binarity}

Our model fits (and hence our results for $f_\mathrm{bol},\teff$, etc.) consider BD~+17~4708 as a single star. However, it is known that BD~+17~4708 shows periodic radial velocity variations with an amplitude of about 4.7~km~s$^{-1}$ over a period of about 220 days (Latham et~al. 1988). Model fits to the radial velocity curve by Latham et~al. suggest a mass function $f(M)=(m_2\sin i)^3/(m_1+m_2)^2=0.0019\pm0.0004M_\odot$, where $m_1$ and $m_2$ are the masses of the stars and $i$ the orbital inclination.

A rough estimate assuming a mass of about $0.9M_\odot$ for the primary (see \S\ref{s:mass}), results in a companion mass of $0.15M_\odot$ (using $<\sin^3i>=3/5$), which corresponds to a late M type star with a $\teff\sim3000$~K. Using Kurucz model fluxes we find that the contribution of the secondary to the UV and blue fluxes, which are much more sensitive to $\teff$ than the IR, is negligible (less than 1/50). Although the secondary contributes a significant flux in the IR (about 1/5), the \textit{shape} of the spectral energy distribution is nearly unchanged. Given that it is this shape along with a scaling factor what determine the best $\teff$ solution for the primary, including the companion flux in the fits will not affect the $\teff$ result significantly. If included, the scaling factor would need to be reduced and the $\teff$ of the primary increased to match the observed UV and blue fluxes. However, given that the orbital inclination is unknown, it is safer to use a single model flux to fit the observed energy distribution, but note that the companion flux may have an important effect in the observed colors, making them redder. This is probably the reason why direct application of the IRFM suggest a lower $\teff$ for the primary compared to the $\teff$ obtained from the flux fit (\S\ref{s:teff}).

The IRFM temperature of this star is 5950~K according to Ram\'{\i}rez \& Mel\'endez (2005a), who used $E(B-V)=0$. Using $E(B-V)=0.01$ the IRFM temperature increases to 6025~K, about 120~K lower than the $\teff$ obtained from the flux fit. It is unlikely that such large difference is due to errors in the absolute infrared flux calibration and/or the zero point determination of the IRFM $\teff$ scale (see Ram\'{\i}rez \& Mel\'endez 2005a for details). The most likely reason for this discrepancy is the flux contributed by the companion, which is more important in the infrared. If the flux at a given wavelength in the infrared ($f_\mathrm{IR}$) is larger then the ratio $R=\fbol/f_\mathrm{IR}$ is smaller compared to that for a single star. This $R$-factor is the $\teff$ indicator in the IRFM, roughly proportional to $\teff^3$ (Ram\'{\i}rez \& Mel\'endez 2005a). Thus, the IRFM temperature obtained for a star with an ignored cool companion is underestimated. In order to account for the 120~K difference (about 2\% error in $\teff$), an error of about 6\% in the $R$-factor is required. In the previous paragraph we estimated a 20\% extra infrared flux due to the companion. Using the same models, the bolometric flux increases by about 10\% if the cool companion is included. If this is the case, the $R$-factor has been underestimated by about 8\%. However, note again that these flux estimates are not accurate due to the large uncertainty in the mass and temperature of the companion.

\section{Spectral line analysis} \label{s:lines}

\subsection{Observations} \label{s:observations}

BD~+17~4708 was observed from McDonald Observatory on October~30, 2004 UT using the 2dcoud\'e spectrograph (Tull et~al. 1995) and the 2.7m Harlan J. Smith telescope. Four individual exposures of 20 min each were obtained at the focal station F3 using grating E2 --a 53.67 gr mm$^{-1}$ R2 echelle from Milton Roy Co.--, a 1.2 arcsec slit, and a $2048\times2048$ Tektronix CCD. The spectra have a FWHM resolving power of $\lambda/\delta \lambda \simeq 60,000$ with full spectral coverage from 3600 \AA\ to 5300 \AA, and substantial but incomplete coverage from 5300 \AA\ to 10000 \AA. The spectra were reduced using the echelle package in IRAF.\footnote{IRAF is distributed by the National Optical Astronomy Observatories, which are operated by the Association of Universities for Research in Astronomy, Inc., under cooperative agreement with the National Science Foundation -- http://iraf.noao.edu} The bias level in the overscan area was modeled with a polynomial and subtracted. An ultra-high signal-to-noise flatfield was used to correct pixel-to-pixel sensitivity variations, and the scattered light was modeled with smooth functions and removed.

The spectra were optimally extracted after cosmic-ray cleanup, and calibrated in wavelength with a Th-Ar hollow cathode lamp (Allende Prieto 2001). By cross-correlating the four individual spectra, we concluded that shifts among them were smaller than 0.2 km s$^{-1}$, and we simply coadded their signal obtaining a single spectrum with a signal-to-noise ratio per pixel in excess of 300 between 5000 \AA\ and 8000 \AA, and in excess of 100 between 4000~\AA\ and 10000 \AA. The individual orders were continuum normalized, combining the signal for the wavelength intervals registered in multiple orders. Similarly to the procedure described by Barklem et~al. (2002), we took advantage of the slow variation  of the blaze function between orders in the normalization process, in order to derive reliable line shapes for the strongest lines.

\subsection{Atomic data}
	
\subsubsection{Iron}

All the $gf$ values for the iron lines used in this work have been measured in the laboratory. No attempts to reduce the line-to-line scatter in the abundances using differential analysis or astrophysical $gf$ values have been made. Thus, our derived iron abundances are strictly given on an \textit{absolute} scale.

The original sources for the transition probabilities of the \ion{Fe}{i} lines are listed by Lambert et~al. (1996), who extensively compared them and concluded that they were all essentially on the same scale, although minor corrections are needed in a few cases. The $gf$ values for the \ion{Fe}{ii} lines have been adopted from Mel\'endez et~al. (2006), who use $gf$ values from theoretical calculations put onto the laboratory scale by means of laboratory lifetimes and branching ratios. Note that these $gf$-values are very similar to those in Lambert et~al. (1996). The mean difference in $\log gf$, in the sense Mel\'endez et~al. 2006 $-$ Lambert et~al. 1996, for 4 \ion{Fe}{ii} lines available in both studies and given in Table \ref{t:linedata}, is only 0.03 dex. A similar comparison with the compilation by Allende Prieto et~al. (2002) reveals that, on average, their $gf$ values are on the same scale as those by Mel\'endez et~al. (2006). However, the line-to-line scatter reduces when adopting the latter set of $gf$ values.

Regarding van der Waals pressure broadening, almost all the damping constants adopted in this work are from Barklem et~al. (2000) and Barklem \& Aspelund-Johansson (2005). For a few \ion{Fe}{i} lines not included in the Barklem et~al. tables, the classical Uns\"old approximation, enhanced by a factor of 2, was adopted. Standard radiative (e.g., Gray 1992) and Stark (e.g., Cowley 1971) broadening approximations, as coded in the latest version of MOOG (Sneden 1973),\footnote{http://verdi.as.utexas.edu/moog.html} were used.

The iron line data and equivalent widths (E.W.s) measured in the spectrum of BD~+17~4708 are given in Table~\ref{t:linedata}. Gaussian profile fitting was used to measure the line E.W.s. 

\begin{table}
\scriptsize \centering
\begin{tabular}{cccrclr} \hline\hline
Wavelength &	Species & E.P.	&	$\log gf$	&	$\log\Gamma$	&	$(1-\alpha)/2$ & E.W. \\	
\AA &	& eV	&	&	rad cm$^3$ s$^{-1}$ &	& m\AA \\	
\hline
4630.120	&	\ion{Fe}{i}	&	2.279	&	$-2.52$	&	$-7.518$	&	0.373	&	7.0	\\
4745.800	&	\ion{Fe}{i}	&	3.654	&	$-1.27$	&	$-7.356$	&	0.300 $\dagger$ 	&	8.6	\\
4939.686	&	\ion{Fe}{i}	&	0.859	&	$-3.34$	&	$-7.748$	&	0.377	&	23.9	\\
4994.129	&	\ion{Fe}{i}	&	0.915	&	$-3.07$	&	$-7.744$	&	0.377	&	28.0	\\
5012.068	&	\ion{Fe}{i}	&	0.859	&	$-2.64$	&	$-7.751$	&	0.377	&	53.0	\\
5044.211	&	\ion{Fe}{i}	&	2.851	&	$-2.03$	&	$-7.280$	&	0.381	&	8.3	\\
5051.634	&	\ion{Fe}{i}	&	0.915	&	$-2.79$	&	$-7.746$	&	0.377	&	42.3	\\
5079.740	&	\ion{Fe}{i}	&	0.990	&	$-3.22$	&	$-7.739$	&	0.378	&	20.1	\\
5110.413	&	\ion{Fe}{i}	&	0.000	&	$-3.76$	&	$-7.826$	&	0.373	&	42.4	\\
5123.720	&	\ion{Fe}{i}	&	1.011	&	$-3.07$	&	$-7.739$	&	0.378	&	26.0	\\
5127.359	&	\ion{Fe}{i}	&	0.915	&	$-3.31$	&	$-7.749$	&	0.377	&	19.9	\\
5150.839	&	\ion{Fe}{i}	&	0.990	&	$-3.00$	&	$-7.742$	&	0.377	&	23.0	\\
5151.911	&	\ion{Fe}{i}	&	1.011	&	$-3.32$	&	$-7.740$	&	0.377	&	17.3	\\
5166.282	&	\ion{Fe}{i}	&	0.000	&	$-4.20$	&	$-7.827$	&	0.373	&	21.2	\\
5171.596	&	\ion{Fe}{i}	&	1.485	&	$-1.78$	&	$-7.688$	&	0.373	&	63.4	\\
5194.941	&	\ion{Fe}{i}	&	1.557	&	$-2.08$	&	$-7.680$	&	0.373	&	45.1	\\
5216.273	&	\ion{Fe}{i}	&	1.608	&	$-2.14$	&	$-7.674$	&	0.372	&	39.4	\\
5227.189	&	\ion{Fe}{i}	&	1.557	&	$-1.23$	&	$-7.681$	&	0.373	&	90.1	\\
5228.376	&	\ion{Fe}{i}	&	4.220	&	$-1.19$	&	$-7.233$	&	0.361	&	4.4	\\
5253.461	&	\ion{Fe}{i}	&	3.283	&	$-1.57$	&	$-7.203$	&	0.386	&	6.2	\\
5307.360	&	\ion{Fe}{i}	&	1.608	&	$-2.98$	&	$-7.678$	&	0.373	&	11.8	\\
5322.041	&	\ion{Fe}{i}	&	2.279	&	$-2.89$	&	$-7.600$	&	0.382	&	4.2	\\
5328.531	&	\ion{Fe}{i}	&	1.557	&	$-1.85$	&	$-7.685$	&	0.374	&	62.0	\\
5332.899	&	\ion{Fe}{i}	&	1.557	&	$-2.78$	&	$-7.685$	&	0.374	&	16.0	\\
5341.023	&	\ion{Fe}{i}	&	1.608	&	$-1.95$	&	$-7.679$	&	0.373	&	52.5	\\
5371.489	&	\ion{Fe}{i}	&	0.958	&	$-1.65$	&	$-7.753$	&	0.376	&	95.9	\\
5373.708	&	\ion{Fe}{i}	&	4.473	&	$-0.74$	&	$-7.123$	&	0.359	&	6.5	\\
5397.127	&	\ion{Fe}{i}	&	0.915	&	$-1.99$	&	$-7.759$	&	0.375	&	81.7	\\
5429.696	&	\ion{Fe}{i}	&	0.958	&	$-1.88$	&	$-7.755$	&	0.376	&	87.8	\\
5432.947	&	\ion{Fe}{i}	&	4.445	&	$-0.94$	&	$-7.153$	&	0.360	&	7.2	\\
5434.523	&	\ion{Fe}{i}	&	1.011	&	$-2.12$	&	$-7.750$	&	0.377	&	72.5	\\
5473.900	&	\ion{Fe}{i}	&	4.154	&	$-0.72$	&	$-7.266$	&	0.380	&	8.4	\\
5497.516	&	\ion{Fe}{i}	&	1.011	&	$-2.85$	&	$-7.752$	&	0.376	&	41.0	\\
5501.465	&	\ion{Fe}{i}	&	0.958	&	$-3.04$	&	$-7.757$	&	0.375	&	32.4	\\
5506.779	&	\ion{Fe}{i}	&	0.990	&	$-2.80$	&	$-7.754$	&	0.376	&	42.1	\\
5543.935	&	\ion{Fe}{i}	&	4.217	&	$-1.04$	&	$-7.263$	&	0.381	&	5.1	\\
5638.262	&	\ion{Fe}{i}	&	4.220	&	$-0.77$	&	$-7.270$	&	0.382	&	6.8	\\
5701.544	&	\ion{Fe}{i}	&	2.559	&	$-2.22$	&	$-7.576$	&	0.382	&	8.5	\\
5905.671	&	\ion{Fe}{i}	&	4.652	&	$-0.69$	&	$-7.144$	&	0.359	&	6.0	\\
5930.179	&	\ion{Fe}{i}	&	4.652	&	$-0.17$	&	$-7.149$	&	0.359	&	24.8	\\
5934.654	&	\ion{Fe}{i}	&	3.928	&	$-1.07$	&	$-7.153$	&	0.377	&	7.6	\\
6003.012	&	\ion{Fe}{i}	&	3.881	&	$-1.06$	&	$-7.181$	&	0.380	&	7.8	\\
6027.050	&	\ion{Fe}{i}	&	4.076	&	$-1.09$	&	$-7.397$	&	0.300 $\dagger$	&	6.3	\\
6056.004	&	\ion{Fe}{i}	&	4.733	&	$-0.40$	&	$-7.130$	&	0.357	&	4.7	\\
6170.507	&	\ion{Fe}{i}	&	4.795	&	$-0.38$	&	$-7.119$	&	0.355	&	6.0	\\
6200.313	&	\ion{Fe}{i}	&	2.608	&	$-2.44$	&	$-7.588$	&	0.382	&	5.0	\\
6213.430	&	\ion{Fe}{i}	&	2.223	&	$-2.48$	&	$-7.691$	&	0.368	&	8.9	\\
6232.641	&	\ion{Fe}{i}	&	3.654	&	$-1.22$	&	$-7.498$	&	0.300 $\dagger$	&	7.9	\\
6265.133	&	\ion{Fe}{i}	&	2.176	&	$-2.55$	&	$-7.699$	&	0.369	&	11.2	\\
6344.148	&	\ion{Fe}{i}	&	2.433	&	$-2.92$	&	$-7.620$	&	0.377	&	3.9	\\
6419.949	&	\ion{Fe}{i}	&	4.733	&	$-0.24$	&	$-7.193$	&	0.363	&	12.0	\\
6609.110	&	\ion{Fe}{i}	&	2.559	&	$-2.69$	&	$-7.610$	&	0.377	&	4.2	\\
6750.152	&	\ion{Fe}{i}	&	2.424	&	$-2.62$	&	$-7.608$	&	0.380	&	6.9	\\
6841.338	&	\ion{Fe}{i}	&	4.607	&	$-0.71$	&	$-7.258$	&	0.367	&	5.7	\\
6855.162	&	\ion{Fe}{i}	&	4.558	&	$-0.74$	&	$-7.347$	&	0.300 $\dagger$	&	5.4	\\
7090.383	&	\ion{Fe}{i}	&	4.230	&	$-1.11$	&	$-7.165$	&	0.376	&	5.0	\\ 
4620.521	&	\ion{Fe}{ii}	&	2.828	&	$-3.21$	&	$-7.878$	&	0.347	&	7.4	\\
4629.339	&	\ion{Fe}{ii}	&	2.807	&	$-2.28$	&	$-7.886$	&	0.372	&	42.0	\\
5197.577	&	\ion{Fe}{ii}	&	3.230	&	$-2.22$	&	$-7.881$	&	0.377	&	32.5	\\
5234.625	&	\ion{Fe}{ii}	&	3.221	&	$-2.18$	&	$-7.881$	&	0.376	&	38.0	\\
5264.812	&	\ion{Fe}{ii}	&	3.230	&	$-3.13$	&	$-7.875$	&	0.350	&	7.1	\\
6432.680	&	\ion{Fe}{ii}	&	2.891	&	$-3.57$	&	$-7.899$	&	0.398	&	5.6	\\
6516.081	&	\ion{Fe}{ii}	&	2.891	&	$-3.31$	&	$-7.899$	&	0.399	&	8.8	\\
\hline
\end{tabular}
\small
\caption{Iron line data. $\Gamma$ and $\alpha$ are the Van der Waals FWHM per perturber at 10,000 K and velocity parameter, respectively. The damping constants for the lines marked with a $\dagger$ correspond to the modified Uns\"old approximation. The last column gives the equivalent widths measured in the spectrum of BD~+17~4708.}
\label{t:linedata}
\end{table}

\subsubsection{Other elements}

For the strong 5180 \AA\ \ion{Mg}{i}\,\textit{b} lines, as well as for the 7774~\AA\ \ion{O}{i} triplet, transition probabilities were obtained from the NIST Atomic Spectra Database.\footnote{http://physics.nist.gov/PhysRefData/ASD/lines\_form.html}

It was difficult to find weak Mg lines with reliable transition probabilities in our spectrum. In fact, we found only the unblended weak 4571~\AA\ \ion{Mg}{i} line, with a reliable $gf$ value from the NIST database. Another weak \ion{Mg}{i} line is that at 5711~\AA, for which we used the solar $gf$ derived by Fuhrmann et~al. (1995).

Data for four weak \ion{Si}{i} lines were taken from the compilation by Allende~Prieto et~al. (2004a), who concentrated on lines with transition probabilities measured in the laboratory or obtained from accurate theoretical calculations.

For the Ca abundance determination, we used the line list by Bensby et~al. (2003) but adopting the $gf$ values from the NIST database instead of using their solar $gf$'s. The Bensby et~al. $gf$ values are systematically lower by about 0.2 dex compared to those obtained from the NIST database. However, the line-to-line scatter in our derived mean Ca abundance is similar for the two sets of $gf$ values.

Radiative, Stark, and van der Waals broadening was computed in the same way as for the iron lines. Note that in this case all lines are weak (with the exception of the \ion{Mg}{i}\,\textit{b} triplet and the 6439~\AA\ \ion{Ca}{i} line) so the use of the modified Uns\"old approximation, when necessary, to obtain the van der Walls damping constants instead of using those from the theory of Barklem et~al. (2000) has no noticeable effect on the abundances.

\begin{table}
\scriptsize \centering
\begin{tabular}{cccrclr} \hline\hline
Wavelength &	Species & E.P.	&	$\log gf$	&	$\log\Gamma$	&	$(1-\alpha)/2$ & E.W. \\	
\AA &	& eV	&	&	rad cm$^3$ s$^{-1}$ &	& m\AA \\	
\hline
7771.944	&	\ion{O}{i}	&	9.146	&	$ 0.37$	&	$-7.469$	&	0.383	&	38.2	\\
7774.166	&	\ion{O}{i}	&	9.146	&	$ 0.22$	&	$-7.469$    &	0.383	&	31.7	\\
7775.388	&	\ion{O}{i}	&	9.146	&	$ 0.00$	&	$-7.469$	&	0.383	&	22.9	\\
4571.096    &  \ion{Mg}{i} &   0.000   &   $-5.39$ &   $-7.645$    &   0.377   &   19.4    \\
5167.321	&  \ion{Mg}{i}	&	2.709	&	$-0.86$	&	$-7.267$	&	0.381	&	----	\\
5172.684	&  \ion{Mg}{i}	&	2.712	&	$-0.38$	&	$-7.267$	&	0.381	&	----	\\
5183.604	&  \ion{Mg}{i}	&	2.717	&	$-0.16$	&	$-7.267$	&	0.381	&	----	\\
5711.100    &  \ion{Mg}{i} &   4.346   &   $-1.67$ &    $-7.218$    &   0.300 $\dagger$	        &   17.6    \\
5708.397    &  \ion{Si}{i} &   4.954   &   $-1.37$ &    $-7.183$    &   0.300 $\dagger$	        &   10.0    \\
5948.540    &  \ion{Si}{i} &   5.082   &   $-1.13$ &    $-7.169$    &   0.300 $\dagger$	        &   12.0    \\
7918.382    &  \ion{Si}{i} &   5.954   &   $-0.51$ &    $-7.010$    &   0.300 $\dagger$	        &   10.3    \\
7932.348    &  \ion{Si}{i} &   5.964   &   $-0.37$ &    $-7.006$    &   0.300 $\dagger$	        &   10.5    \\
4526.928    &  \ion{Ca}{i} &   2.709	&   $-0.42$ &   $-7.021$     &   0.300 $\dagger$	        &   15.8    \\
4578.551    &  \ion{Ca}{i} &  	2.521	&   $-0.56$ &   $-7.125$     &   0.300 $\dagger$	        &   18.3    \\
5512.980    &  \ion{Ca}{i} &  	2.933	&   $-0.30$ &   $-7.269$    &   0.300 $\dagger$	        &   16.7    \\
6166.439    &  \ion{Ca}{i} &  	2.521	&   $-0.90$ &   $-7.146$    &   0.372   &   11.3    \\
6169.042    &  \ion{Ca}{i} &  	2.523	&   $-0.54$ &   $-7.146$    &   0.372   &   21.0    \\
6169.563    &  \ion{Ca}{i} &  	2.526	&   $-0.27$ &   $-7.145$    &   0.372   &   30.6    \\
6439.075    &  \ion{Ca}{i} &  	2.526	&   $ 0.47$ &   $-7.569$    &   0.379   &   69.4    \\
6471.662    &  \ion{Ca}{i} &  	2.526	&   $-0.59$ &   $-7.570$    &   0.380   &   24.4    \\
6493.781    &  \ion{Ca}{i} &  	2.521	&   $ 0.14$ &   $-7.571$    &   0.381   &   54.8    \\
6499.650    &  \ion{Ca}{i} &  	2.523	&   $-0.59$ &   $-7.571$    &   0.381   &   18.1    \\
\hline
\end{tabular}
\small
\caption{As in Table \ref{t:linedata} for the O, Mg, Si, and Ca lines. Equivalent widths are given for all but the strong \ion{Mg}{i}\,\textit{b} lines.}
\label{t:linedata2}
\end{table}

\subsection{Modeling}

Spectrum synthesis was performed using MOOG (Sneden 1973) and the non-LTE codes TLUSTY and SYNSPEC (Hubeny 1988, Hubeny \& Lanz 1995).\footnote{http://nova.astro.umd.edu} For practical reasons, MOOG was preferred for matching the equivalent widths of the iron lines, while SYNSPEC was used to fit the profiles of strong lines. The same scaled solar abundances (those by Grevesse \& Sauval 1998) were used in the two codes and thus, only very small differences, mainly due to the continuum opacity calculations, may be present when comparing the results from the two codes. All line formation calculations were done assuming LTE, with the only exception of the 7774~\AA\ triplet (see \S\ref{s:oxygen}).

For the line-profile fitting, all the synthetic profiles have been broadened by convolving the theoretical spectra with Gaussian profiles of $\mathrm{FWHM}=0.18$ \AA\ in the red and $\mathrm{FWHM}=0.21$ \AA\ in the near infrared. These FWHM values are empirically determined global broadening parameters that fit very well weak lines with reliable atomic data. In fact, given the resolution $R\sim60,000$ ($\delta\lambda\sim0.09~\AA$ at 5500 \AA, $\delta\lambda\sim0.13~\AA$ at 7780 \AA), a solar-like macroturbulent velocity of 1.5~km~s$^{-1}$ ($\delta\lambda\sim0.03$ \AA\ at 5500 \AA, $\delta\lambda\sim0.04$ \AA\ at 7780 \AA), and a low projected rotational velocity of $v\sin i\sim3$~km~s$^{-1}$, our estimates for the FWHM values are well justified.

We used the most recent Kurucz no-overshoot model atmospheres with $\alpha$-element enhancement (e.g., Kurucz 1970, 1979). The use of models with the convective overshooting option switched on produces an almost constant shift of less than 0.1 dex in the abundance scale but preserves abundance ratios as well as the difference in the mean Fe abundances from \ion{Fe}{i} and \ion{Fe}{ii} lines reported in \S\ref{s:iron}. Use of models with solar scaled abundances (i.e., without $\alpha$-element enhancement) produced essentially the same abundances. For each set of atmospheric parameters adopted, a microturbulent velocity $v_t$ was derived by making the abundances from the \ion{Fe}{i} lines independent of their reduced equivalent widths (E.W.$/\lambda$).

\subsection{$\teff$ from the Balmer lines} \label{s:balmer}

\begin{figure}
 \centering
 \includegraphics[width=7.5cm]{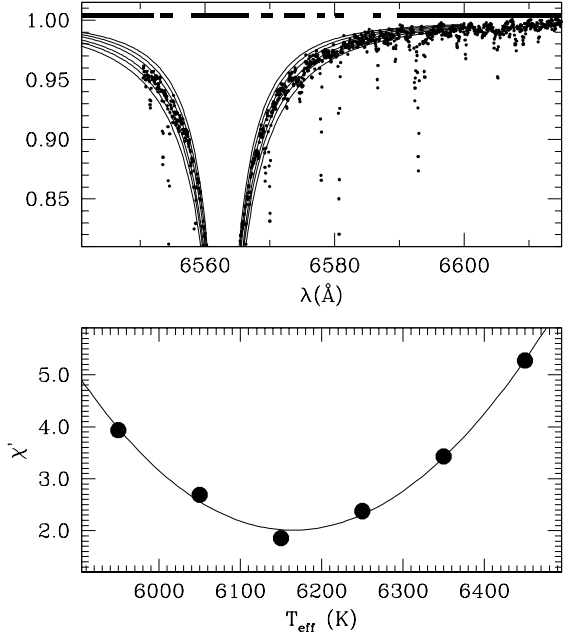}
 \caption{Top panel: the observed H$\alpha$ profile (points) is shown along with model line profiles of $\teff=5950,6050,6150,6250,6350,6450$~K (solid lines). The bars on top show the regions excluded from the $\chi^2$ calculation. Bottom panel: quality of the fits in the top panel, as measured by $\chi'=\sqrt{\chi^2/(n-1)}$ where $\chi^2=\sum_{i=1}^n[(f_\mathrm{obs}-f_\mathrm{kur})_i^2/\sigma_i^2]$ ($\sigma$ is the error in $f_\mathrm{obs}$ only), as a function of $\teff$ (filled circles). The solid line is a cubic fit to the filled circles.}
 \label{f:balmers}
\end{figure}

Balmer line-profiles were synthesized using the prescription by Barklem et~al. (2002) but adopting Kurucz model atmospheres. In short, Stark broadening was computed according to Stehl\'e \& Hutcheon (1999) while self-broadening is from Barklem et~al. (2000). The shapes of strong lines like H$\alpha$ are very well determined in our spectrum by fitting the blaze shape of each clean order (those free from very strong line absorption), and modeling the smooth variation of the shape of the blaze with order number to set the continuum (see \S2 in Barklem et~al. 2002 for details).

The results for H$\alpha$ are shown in Fig.~\ref{f:balmers}, along with the observed profile in the spectrum of BD~+17~4708. As shown in Fig.~\ref{f:balmers}, the wings of the H$\alpha$ line are very sensitive to $\teff$. Unfortunately, due to contamination by metallic lines, the H$\beta$ profile is difficult to use as a temperature indicator.

Excluding the most prominent metallic features, as shown in Fig.~\ref{f:balmers}, a $\chi^2$ test favors a $\teff=6165$~K from H$\alpha$. For the $\chi^2$ calculations, the H$\alpha$ line was cut at 6550~\AA\ since shorter wavelengths fall outside the CCD. In fact, our spectrum in this order goes down to about 6535 \AA, but the data points between 6535 \AA\ and 6550 \AA\ are, observationally, somewhat more uncertain. If the blue wing of the H$\alpha$ line is ignored altogether (to minimize further the observational errors), the temperature increases only by 20~K. On the other hand, the sensitivity to $\teff$ decreases for $\lambda>6590$ \AA\ and, given the S/N, introducing these longer wavelengths to assess the quality of the fits increases only the absolute $\chi^2$ values without changing significantly the inferred $\teff$.

Our H$\alpha$ temperature is in excellent agreement with that given by Asplund et~al. (2006), who obtained $\teff=6183$~K from fits to their H$\alpha$ profile using MARCS models but the same treatment for the line broadening. Both, ours and Asplund et al. H$\alpha$ temperatures are in good agreement with the $\teff$ derived from the fitting of the spectral energy distribution.

\subsection{Surface gravity from strong lines} \label{s:gravity}

The stellar surface gravity of a star can, in principle, be obtained from an estimate of its mass and its measured trigonometric parallax, besides reasonable estimates of $\teff$ and $\feh$. The mass of a nearby star can be reasonably estimated from its position on a color-magnitude diagram using theoretical isochrones but the \textit{Hipparcos} parallax of stars farther than 100~pc, as is the case for BD~+17~4708, is quite uncertain and therefore their trigonometric $\logg$ values are not reliable. In fact, using this method we only obtain a weak constrain: $3.8<\logg<4.6$.

\begin{figure*}
 \centering
 \includegraphics[width=12.5cm]{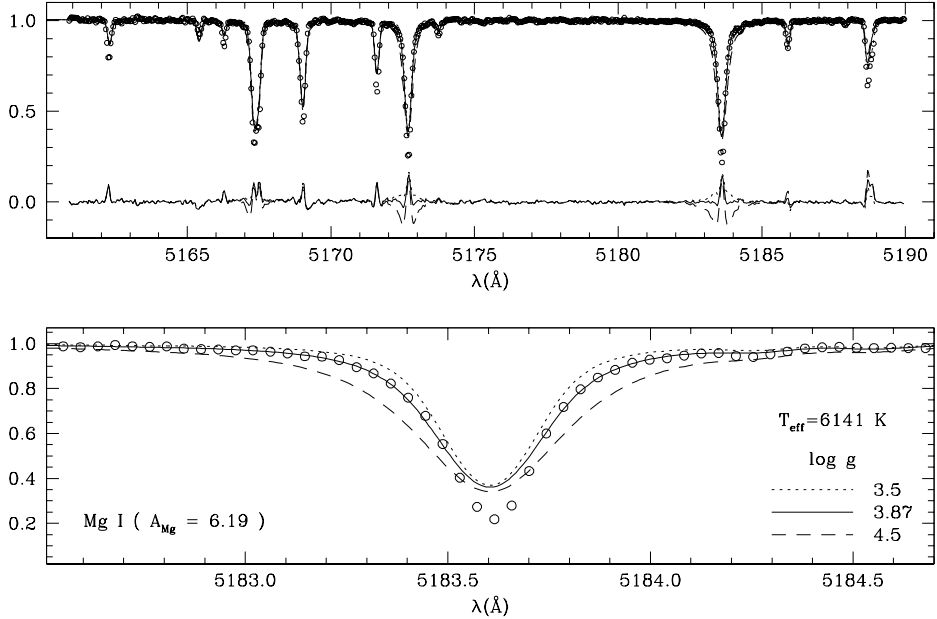}
 \caption{Top panel: the region around the strong \ion{Mg}{I}\,\textit{b} triplet as given by the observations (open circles), and as predicted by models of $\teff=6141$~K, $\feh=-1.74$, $A_\mathrm{Mg}=6.19$, and three different values of the surface gravity ($\logg=3.5,3.87,4.5$). Residuals of the fit are also shown. Bottom panel: zoom into the 5183.6 \AA\ line.}
 \label{f:s5185hot}
\end{figure*}

Fortunately, the wings of some strong lines are sensitive to the $\logg$ value and are less affected by $\teff$ and can thus be used to constrain the surface gravity. In the BD~+17~4708 spectrum, only the two strongest lines of the \ion{Mg}{i}\,\textit{b} triplet at 5172.7 \AA\ and 5183.6 \AA\ seem suitable for this kind of analysis (Fig.~\ref{f:s5185hot}). Note that the cores of the strong lines are strongly affected by non-LTE and we do not expect good fits in the line centers, only the wings should be used to assess the quality of the fits.

Fig.~\ref{f:s5185hot} shows that an excellent fit to the wings of the \ion{Mg}{i}\,\textit{b} triplet is obtained with $\logg=3.87$ when the Mg abundance is set to $A_\mathrm{Mg}=6.19$, as derived in \S\ref{s:alpha}. Our preferred solution for the surface gravity is thus $\logg=3.87\pm0.08$. The error bar was estimated by propagating the error of 0.06~dex in $A_\mathrm{Mg}$, which includes the 50~K error in $\teff$.

The \ion{Mg}{i}\,\textit{b} triplet is very strong and contaminated by metallic lines in the solar spectrum. Nevertheless, adopting the same procedure we used to obtain the $\logg$ value of BD~+17~4708, we were able to satisfactorily reproduce the wings of these lines in the solar spectrum of Kurucz et~al. (1984) with the standard solar $\logg=4.44$ and Mg abundance of $A_\mathrm{Mg}=7.53$ (Asplund et~al. 2005). A visual inspection showed that the accuracy of our method of $\logg$ determination in the Sun is about 0.1~dex. Therefore, the $\logg$ value we derive for BD~+17~4708 is accurate at the 0.1~dex level in the absolute scale.

\subsubsection{Mass, age, and radius from theoretical isochrones} \label{s:mass}

Isochrones in the theoretical HR diagram ($\teff$ vs. $\logg$) instead of the observational HR diagram (absolute magnitude vs. color) can be used to estimate the mass and age of a star if its parameters, but not necessarily its distance, are known with accuracy. This is the case of BD~+17~4708.

Although the mass estimates are normally accurate using this approach, the age determinations may be subject to severe systematic errors and statistical biases (see, e.g., Pont \& Eyer 2004) so they must not be considered accurate even if the stellar parameters are. We used the Bertelli et~al. (2004) isochrones, as in Allende Prieto et~al. (2004a), to estimate the mass ($M$) and age ($t$) of BD~+17~4708. The Bertelli et~al. isochrones were computed using solar-scaled chemical compositions. In metal-poor stars, however, the $\alpha$-element enhancement ($[\alpha/\mathrm{Fe}]\simeq+0.4$ in our case, \S\ref{s:alpha}) has an important effect on these calculations (e.g., VandenBerg et~al. 2000, Kim et~al. 2002) although its effect on the mass and age derived from the isochrones is relatively small (about $+0.1M_\odot$ and $-0.4$~Gyr in our case). Therefore, we increased the $\feh$ value of BD~+17~4708 by about 0.2~dex to mimic the $\alpha$-element enhancement, as suggested by Salaris et~al. (1993) and obtained $M=0.91^{+0.11}_{-0.04}M_\odot$ and $t=8.8^{+2.6}_{-1.8}$~Gyr ($2\sigma$ errors).

The referee noted that using our derived parameters the Vandenberg et al. (2000) isochrones suggest an age close to 10 Gyr but if the $\logg$ value is increased to 4.05 then the age would increase to about 13.5 Gyr. Using the Bertelli et~al. isochrones and $\logg=4.05$, we obtained $t=11.4^{+1.3}_{-4.2}$~Gyr, i.e., an increase of 2.6~Gyr in the mean age. The halo is believed to have an age of about 13 Gyr (e.g., Schuster et~al. 2006). However, given that this mean age is calculated using large samples of halo stars and in some cases sophisticated statistics, this should not be used to discard or confirm ages of individual stars. Note that, for example, our derived age is in agreement with that given by Nordstr\"om et~al. (2004), who took into account the statistical biases in isochrone age determinations described in Pont \& Eyer (2004). Also, some halo stars, even more metal-poor than BD~+17~4708, seem to be younger than the mean age of the halo (see, e.g., Table 2 in Li \& Zhao 2004, who give a compilation of radioactive ages, including theirs).

The radius that we obtain using isochrones is about $1.8R_\odot$, with a 2-$\sigma$ range that goes from 1.5 to $2.3R_\odot$ if we include systematic errors in our $\teff$ and $\logg$ estimates. Although inaccurate, the \textit{Hipparcos} parallax constrains the radius to a 1-$\sigma$ range from 1.1 to 1.6 $R_\odot$, if we adopt our angular diameter (\S\ref{s:theta}). A slightly higher $\logg$ value, for example $\logg=4.0$, would result in $R\simeq1.4R_\odot$. Note that this would still be in good, albeit marginal, agreement with our result for $\logg$ considering the random error bar (0.08~dex) and a possible systematic error in the absolute scale (about 0.10~dex).

\subsection{The iron abundance} \label{s:iron}

The customary approach to determination of the Fe abundance invokes LTE for the excitation and ionization of iron neutral atoms and singly-charged ions. An estimate of the effective temperature is obtained by the requirement that the derived Fe abundances from the \ion{Fe}{i} lines be independent of their excitation potential. Application of this requirement generally demands a prior determination of the microturbulence ($v_t$), often from the same set of \ion{Fe}{i} lines and the condition that the Fe abundance be independent of a line's E.W. Then, the imposition of ionization equilibrium through the requirement that the \ion{Fe}{i} and \ion{Fe}{ii} lines give the same Fe abundance defines a locus in the ($\teff,\logg$) plane which with the $\teff$ from the \ion{Fe}{i} lines (or another source) serves to determine the surface gravity.

\begin{figure*}
 \centering
 \includegraphics[width=12cm]{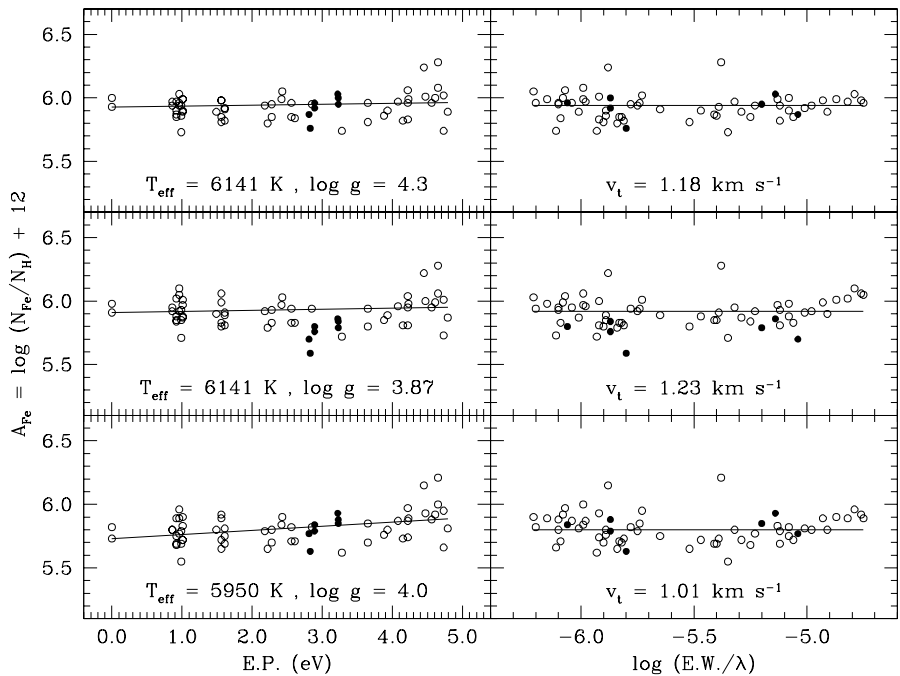}
 \caption{Abundance of iron from \ion{Fe}{i} (open circles) and \ion{Fe}{ii} lines (filled circles), as determined using different combinations of $\teff$, $\logg$, and $v_t$; as a function of excitation potential and reduced equivalent width. The solid lines are linear fits to the \ion{Fe}{i} data only.}
 \label{f:feabund}
\end{figure*}

\begin{figure}
 \centering
 \includegraphics[width=6.9cm]{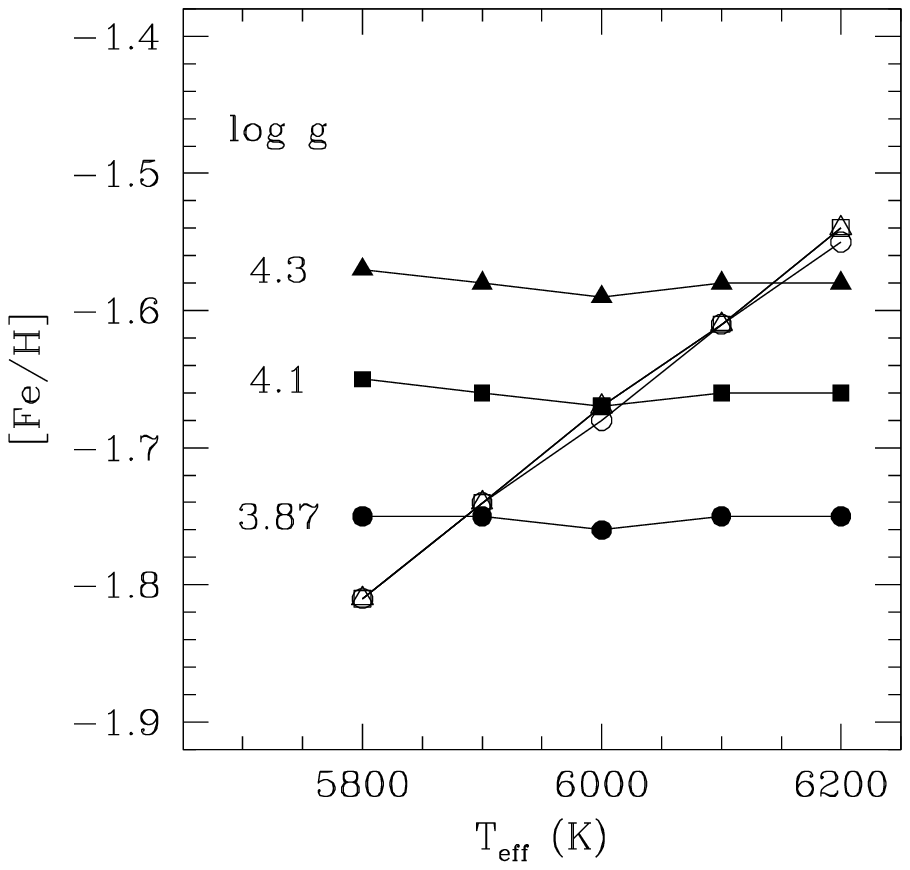}
 \caption{Mean abundance of iron from \ion{Fe}{i} (open symbols) and \ion{Fe}{ii} (filled symbols) lines as a function of $\teff$ and $\logg$. Here $\feh=\mathrm{A}_\mathrm{Fe}-7.51$, i.e., we used A$_\mathrm{Fe}=7.51$ for the Sun.}
 \label{f:iron}
\end{figure}

We used 56 \ion{Fe}{i} lines covering the excitation potential (E.P.) range from 0 to 5 eV and 7 \ion{Fe}{ii} lines to derive the iron abundance for various choices of atmospheric parameters (Figs.~\ref{f:feabund} and \ref{f:iron}). The lines we selected have E.W. between 4 and 100~m\AA\ to avoid errors due to noise and saturation.

The \ion{Fe}{i} lines with an almost 5 eV range in excitation potential demand an effective temperature only slightly larger than about 6141~K (Fig.~\ref{f:feabund}), almost independently of the choice of surface gravity, and confirming the temperature provided by the flux distribution. A temperature of 5950~K is demonstrably too low.

A correlation between the excitation potential of the lines and their reduced equivalent widths may lead to degenerate solutions for the ($\teff,v_t$) pair. There is no such correlation for our \ion{Fe}{i} lines with $\mathrm{E.P.}>2$ eV. Our \ion{Fe}{i} lines with $\mathrm{E.P.}<2$ eV do not show such correlation either but they are all stronger (i.e., they all have larger reduced equivalent widths) than those with $\mathrm{E.P.}>2$ eV. The strong, low E.P. lines allow to better determine $v_t$. Notice, however, that the $A_\mathrm{Fe}$ vs. E.P. relations shown in Fig.~\ref{f:feabund} do not change dramatically if the lowest E.P. lines are avoided.

\begin{figure}
 \centering
 \includegraphics[width=8.5cm]{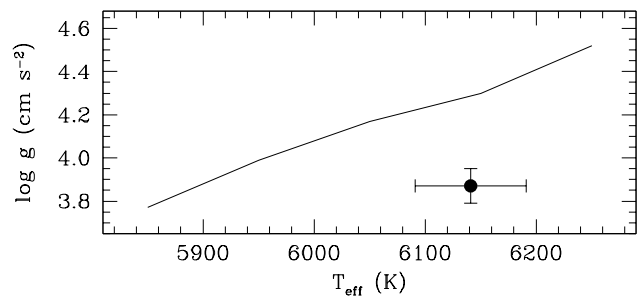}
 \caption{Locus of the $\teff$ and $\logg$ values for which ionization equilibrium of Fe lines is satisfied. The circle with the error bars correspond to our best estimates of $\teff$ from the flux distribution and $\logg$ from the fit of the \ion{Mg}{i}\,\textit{b} lines.}
 \label{f:locus}
\end{figure}

The LTE ionization equilibrium is satisfied at the locus shown in Fig.~\ref{f:locus}. At $\teff=6141$~K, the locus (see also Fig.~\ref{f:iron}) corresponds to about $\logg=4.3$, a value higher than that provided by the fit to the \ion{Mg}{i}\,\textit{b} lines. Note that this result is inconsistent with that found by Edvardsson (1988), who concludes that the strong line gravities are larger than those obtained from the ionization balance in a sample of subgiants with metallicities higher than about $\feh=-0.5$. At that lower value of $\logg=3.87$ and $\teff=6141$~K, the Fe abundance from \ion{Fe}{ii} lines is about 0.15 dex less than that from the \ion{Fe}{i} lines, as it is clearly seen in the middle panels of Fig.~\ref{f:feabund}. Ionization equilibrium at $\logg=3.87$ is achieved if $\teff\simeq5900$~K, but this temperature is judged to be too low.

Our analysis shows that a consistent analysis of the flux distribution and the \ion{Mg}{i}\,\textit{b}, \ion{Fe}{i}, and \ion{Fe}{ii} lines cannot be found within the constraints of a classical LTE model atmosphere analysis. The inconsistencies, almost certainly, cannot be ascribed to the accumulation of errors in the flux and line data. One must suspect a failure of the classical atmosphere and/or the breakdown of the LTE assumption.

Introduction of departures from LTE into the formation of iron lines within a classical model atmosphere constructed assuming LTE for all sources of continuous and line opacity calls for atomic data on radiative and collisional processes far beyond the restricted need for the corresponding LTE analysis. The main non-LTE effect on the Fe lines has been shown to be an overionization of neutral Fe atoms resulting from the UV flux (e.g., Athay \& Lites 1972). This effect is the more severe for metal-poor stars owing, principally, to the reduced line blocking in the UV.

Calculations reported for HD~140283, a star more metal-poor and cooler than BD~+17~4708, show that, using 1D model atmospheres (see next paragraph), the Fe abundance from the \ion{Fe}{i} lines might be increased by up to about 0.5 dex for a non-LTE analysis (e.g., Korn et~al. 2003, Shchukina et~al. 2005) while leaving the Fe abundance from the \ion{Fe}{ii} lines nearly unchanged. If these non-LTE effects are taken into account in our case, they would increase further the difference between the Fe abundance from the neutral and ionized lines and would require even higher surface gravities to achieve ionization equilibrium. Note, however, that the role of inelastic collisions with neutral hydrogen in the non-LTE calculations needs to be explored in more detail given that they may significantly reduce the size of the non-LTE corrections to the \ion{Fe}{i} abundance (Korn et~al. 2003). In fact, we find that they are very important for the oxygen abundance determination from the 7774~\AA\ triplet although the formulation commonly adopted (that by Steenbock \& Holweger 1984) is questionable and the resulting abundances uncertain (see \S\ref{s:oxygen}).

Classical atmospheres with their assumption of plane parallel homogeneous layers in hydrostatic equilibrium cannot represent the stellar granulation resulting from convective instabilities. Modeling of granulation in main-sequence stars including metal-poor examples is beginning with attendant analyses of LTE and non-LTE line formation (Asplund 2005). The latter models are commonly referred to as 3D models with classical models as 1D models. Calculations for HD~140283 suggest that the non-LTE Fe abundance from the \ion{Fe}{ii} lines is increased by about 0.3 dex in going from a 1D to a 3D model of the same atmospheric parameters but the non-LTE abundance from \ion{Fe}{i} lines is unchanged (Shchukina et~al. 2005). 

Taking the results for HD~140283 at face value, and applying them to our case, the switch from the 1D LTE analysis to the 3D non-LTE one (without H collisions) would lead to an increase of about $+0.2$ dex in the $\ion{Fe}{i}-\ion{Fe}{ii}$ abundance difference (the \ion{Fe}{i} abundance increasing by 0.5 dex due to non-LTE in 1D and the \ion{Fe}{ii} abundance increasing by 0.3 dex due to 3D effects in non-LTE). Thus, if the sense and approximate magnitude of these effects applies to BD~+17~4708, the disagreement between the surface gravity derived from the LTE analysis of the \ion{Mg}{i}\,\textit{b} lines and Fe ionization equilibrium is increased. Excitation by collisions with H atoms could alleviate the disagreement discussed somewhat but no reliable theory to include them in the non-LTE calculations is available at present (see \S\ref{s:oxygen}).

There is clearly a need for a fuller exploration of the non-LTE effects (in particular H collisions) both in the construction of the model atmosphere and in the line formation. Detailed testing of 3D models is a necessity with confrontation between predictions of the energy distribution, line strengths, wavelengths, and asymmetries. Pending this major challenge, we conclude that the best fits to the data so far have been achieved with $\teff=6141$~K and $\logg=3.87$. Given that the \ion{Fe}{ii} lines seem to be less affected by errors in $\teff$ or non-LTE (see, e.g., Th\'evenin \& Idiart 1999, but see also Shchukina et~al. 2005), we disregard the ionization balance condition and adopt the mean abundance from the \ion{Fe}{ii} lines only as our $\feh$ indicator. Thus, our preferred solution for the metallicity of BD~+17~4708 is $\feh=-1.74\pm0.09$, which, for our inferred solar iron abundance,\footnote{Our line list and adopted atomic data result in $A_{\mathrm{Fe},\odot}=7.51\pm0.08$, with no significant difference between the \ion{Fe}{i} and \ion{Fe}{ii} abundances, although not all the lines could be used in the solar spectrum due to saturation. Details will be given in Ram\'{\i}rez et~al. (2006).} corresponds to $A_\mathrm{Fe}=5.77\pm0.09$.\footnote{Unless otherwise noted, all the error bars for the abundances are 1-$\sigma$ errors.}

\subsection{Mg, Si, and Ca abundances} \label{s:alpha}

Besides their importance in the study of Galactic chemical evolution, the abundances of these elements are relevant for stellar structure and evolution calculations (e.g., VandenBerg et~al. 2000). Although they are less abundant than carbon, nitrogen, and oxygen, they are important sources of opacity at high temperatures.

It is well established that the [$\alpha$/Fe] ratios, where $\alpha$ is an $\alpha$ element (e.g., Mg, Si, Ca, etc.), in the majority of metal-poor stars are well above solar. Most authors agree on [$\alpha$/Fe] ratios of about $+0.3\pm0.1$~dex in the halo (e.g., Carretta et~al. 2000, Gratton et~al. 2000, Idiart \& Th\'evenin 2000, Arnone et~al. 2005, Barklem et~al. 2005).

The Mg abundance was obtained by averaging the abundance derived from the 4571~\AA\ and 5711~\AA\ lines. However, due to the more reliable transition probability of the former, we gave a double weight to the Mg abundance obtained from the 4571\AA\ line. In this way we find $A_\mathrm{Mg}=6.19\pm0.05$. The four Si lines listed in Table~\ref{t:linedata2} resulted in $A_\mathrm{Si}=6.12\pm0.06$ while the Ca lines suggest $A_\mathrm{Ca}=4.93\pm0.06$.

Almost all the Mg, Si, and Ca lines used to derive the abundances given above are very strong ($\mathrm{E.W.}>\sim90$ m\AA) in the solar spectrum, which makes them unsuitable to derive solar abundances due to saturation effects. Instead, we used the solar abundances derived by Asplund et al. (2005) to obtain $\mathrm{[Mg/Fe]}=0.40\pm0.10$, $\mathrm{[Si/Fe]}=0.35\pm0.11$, and $\mathrm{[Ca/Fe]}=0.36\pm0.11$, where the error due to the uncertainties in $\feh$ and $\teff$ have been included. The mean $\mathrm{[\alpha/Fe]}$ ratio is $0.37\pm0.06$, where the error bar here is a standard error.

\section{The \ion{O}{i} triplet: non-LTE effects and [O/Fe] ratio} \label{s:oxygen}

The observed oxygen abundances are relevant for studies of chemical evolution of the Galaxy and supernovae yields (e.g., Wheeler et~al. 1989), as well as for the modeling of stellar structure and evolution (e.g., VandenBerg \& Stetson 1991).

The IR triplet lines are strong enough as to be detected in most metal-poor FGK dwarfs and it has been long known that they suffer from strong departures from LTE (e.g., Eriksson \& Toft 1979, Kiselman 1993, Shchukina et~al. 2005). Accurate oxygen model atoms are available in the literature (e.g., Allende~Prieto et~al. 2003a, Shchukina et~al. 2005), which allow to confidently perform non-LTE calculations. The importance of inelastic collisions with neutral H, often neglected in the non-LTE calculations, have been explored by Allende~Prieto et~al. (2004b) and shown to be necessary to accurately reproduce the center-to-limb variation of the triplet line-profiles in the solar spectrum. Studies of the \ion{O}{i} triplet are also important because the triplet lines are excellent probes of the physics of line formation (e.g., Reetz 1999).

\begin{figure}
 \centering
 \includegraphics[width=7.5cm]{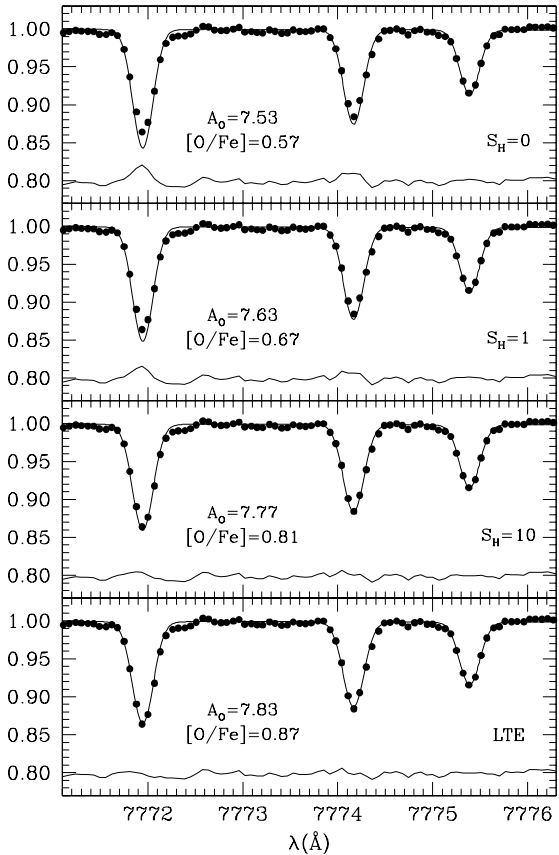}
 \caption{Observed profile of the \ion{O}{i} triplet in BD~+17~4708 (filled circles). Non-LTE model fits to the data (solid lines) are shown for $S_H=0,1$, and 10. LTE fits are also shown. The abundances adopted in each case, given in each panel, have been chosen to fit the reddest line of the triplet. Residuals, shifted by +0.80, are show at the bottom of each panel.}
 \label{f:tri1}
\end{figure}

Using the exact same atomic model and non-LTE calculations as those used in Allende~Prieto et~al. (2004b),\footnote{Allende Prieto et al. used both 1D and 3D model atmospheres in their study. We remind the reader that our work is restricted to 1D Kurucz models.} we computed non-LTE line-profiles and compared them to those observed in the spectra of BD~+17~4708 (Fig.~\ref{f:tri1}). LTE profiles were also computed for completeness. Non-LTE calculations were made both with and without including H collisions. In the former case, the simple approximation formula by Drawin (1968), enhanced by an empirical factor $S_H$, as suggested by Steenbock \& Holweger (1984), was adopted.

The effect of including H collisions is evident from the fits in Fig.~\ref{f:tri1}. Our non-LTE calculations without H collisions, when forced to fit the reddest line of the triplet by tuning the oxygen abundance, overestimate the two strongest lines of the triplet. As $S_H$ is increased, the three lines are better fitted, simultaneously, while reducing the size of the non-LTE correction to the oxygen abundance at the same time. A very good fit to the BD~+17~4708 triplet profile is found with $S_H=10$. Note, however, that in the solar case the $S_H=1$ model fits accurately the center-to-limb observations (Allende~Prieto et~al. 2004b). Although Allende Prieto et~al. did not test non-LTE calculations with $S_H=10$, such case would lead to an oxygen abundance in disagreement with the other oxygen abundance indicators.

The triplet lines originate from the 3s$^5$S$^0$ to 3p$^5$P transition. The upper level consists of three states with very similar energies but noticeably different transition probabilities. This implies that the depths of formation of the three lines are different. The bluest, strongest line is formed in an upper layer while the reddest, weakest line of the triplet is formed in a deeper layer. Since the density of neutral hydrogen decreases with depth in the stellar atmosphere, the collisional rates due to neutral H are more important in the upper layers thus thermalizing more efficiently the 3p$^5$P levels, which results in a weakening of the line strengths. Considering the different formation depths of the three lines and the fact that the 3p$^5$P level becomes progressively more thermalized with height due to H-collisions, the bluest line of the triplet gets more weakened than the reddest line. This reasoning is consistent with what is shown in Fig.~\ref{f:tri1}.

Adopting $A_\mathrm{O}=8.70$ for the Sun, we find an LTE [O/Fe] ratio of 0.87 for BD~+17~4708. The best non-LTE fit to the data, that for $S_H=10$, reduce this ratio to 0.81. Oxygen abundances in metal-poor stars have been determined by several groups using different lines and types of analyses without general agreement. The so-called `oxygen abundance problem' is complex and still open (see, e.g., Nissen et~al. 2002, Fulbright \& Johnson 2003, Mel\'endez et~al. 2006). Discrepancies regarding whether the [O/Fe] ratios remain constant at about +0.5 for the most metal-poor stars or if they increase as lower $\feh$ values are reached still exist although most authors favor constant [O/Fe] ratios. In view of our results for BD~+17~4708, we conclude that H-collisions are an important ingredient in the non-LTE computations of the triplet, not to be ignored in metal-poor stars. However, it is still unclear if the approximation adopted in this work is accurate, given that two different values of $S_H$ are needed to fit the spectra of the Sun and a metal-poor star. In any case, when non-LTE oxygen abundances are inferred from the triplet, the three observed line profiles must be accurately reproduced, simultaneously, by the models. At present, our analysis does not allow us to give a reliable estimate of the oxygen abundance of BD~+17~4708.

\section{Conclusions}

The high accuracy with which the spectral energy distribution of the SDSS standard BD~+17~4708 has been measured has allowed us to provide a reliable estimate of its effective temperature. We have then used spectral line analysis to infer consistent parameters for this star.

Once the degeneracy between $\teff$ and $E(B-V)$ in the model fits to the observed flux distribution is broken by independent estimates of $E(B-V)$, which included a detailed modeling of interstellar absorption features in the observed spectrum, we obtain the following parameters, given here along with reasonable estimates of the ($1\sigma$) error bars: $\teff=6141\pm50$~K, $\feh=-1.74\pm0.09$, $\logg=3.87\pm0.08$, $E(B-V)=0.010\pm0.003$. The spectral energy distribution also allowed us to obtain reliable values for the bolometric flux, $f_\mathrm{bol}=4.89\pm0.10\times10^{-9}$~erg~cm$^{-2}$~s$^{-1}$, and angular diameter, $\theta=0.1016\pm0.0023$~mas, of the star. We thus provide accurate (in an absolute sense) parameters for the spectrophotometric standard of the \textit{Sloan Digital Sky Survey}.

Compared to previous spectroscopic studies, our $\teff$ is higher by about 190~K, which has a severe impact on classical abundance analyses of moderately metal-poor stars. For example, this increase ruins the good agreement between the iron abundance derived from \ion{Fe}{i} and \ion{Fe}{ii} lines found when a lower $\teff$ is used. Despite this, all other features on the spectrum (e.g., the Balmer lines or the strong Mg lines used to constrain the $\logg$ value) seem to be more consistent with the models when a high $\teff$ is adopted. In particular, the excitation balance of \ion{Fe}{i} lines is satisfied with the high $\teff$ but not with the lower value.

We also determine the mean abundance of $\alpha$-elements ($\mathrm{[\alpha/Fe]}=0.37\pm0.06$). The non-LTE modeling of the permitted oxygen triplet lines should include the effect of collisions with neutral H to reasonably reproduce the observations but a better physical treatment of H collisions is needed.

\begin{acknowledgements}
This work was supported in part by the Robert~A.~Welch Foundation of Houston, Texas. CAP research is funded by NASA (NAG5-13057 and NAG5-13147). SR would like to acknowledge support provided by NASA through Hubble Fellowship grant HF-01190.01-A awarded by the Space Telescope Science Institute, which is operated by the Association of Universities for Research in Astronomy, Inc., for NASA, under contract NAS5-26555. We thank Jorge~Mel\'endez for useful advice on interstellar reddening from maps, Martin Asplund for sending us a MARCS model atmosphere, and the referee, Andreas Korn, for comments and suggestions that helped to improve the paper.
\end{acknowledgements}

\appendix

\section{Comparison with the literature} \label{s:literature}

Here we compare the atmospheric parameters derived in this paper for BD~+17~4708 with those given in the literature (Table~\ref{t:literature}), with the exception of Peterson (1981), which is not a CCD-based paper. For the absolute iron abundance ($\afe$) comparison, corrections due to the different solar iron abundances adopted in each study and effective temperature difference effects (\ion{Fe}{i} lines only) are taken into account. For the latter we adopt a correction of 0.2 dex per 300 K ($6.67\times10^{-4}$ dex K$^{-1}$, according to Fig.~\ref{f:iron}) when necessary. In each case, `our' $\afe$ value refers to the abundance we would derive with the parameters adopted by each author or group.

\begin{itemize}

\item \textit{Rebolo et~al. (1988)} adopted a temperature scale similar to that used by Peterson (1981), which is coupled to older Kurucz models by photometric calibrations based essentially on synthetic photometry, but with different color calibrations. They derive $\teff=5890$~K. Their $\logg=4.0$ was obtained from photometric calibrations based on Str\"omgren photometry. Only one \ion{Fe}{i} line was used to derive their $\afe=5.80$. We confirmed the accuracy of the E.W. given in their paper and for this line only we derive the exact same abundance when using the Rebolo et~al. parameters.

\item \textit{Magain (1989)} used a photometric calibration based on the IRFM (Magain 1987) to derive $\teff=5960$~K. This relatively low IRFM $\teff$ is due to the assumption of zero reddening. The $\logg=3.40$ derived in this work is too low and it does not satisfy ionization equilibrium with our line data when the low $\teff$ is adopted. Some, but not all, the $gf$ values used in this work are based on a solar analysis. They derive $\afe=5.75$, which is about 0.05 dex lower than our result for the \ion{Fe}{i} lines.

\item \textit{Axer et~al. (1994)} inferred their $\teff=6100$~K from model fitting of the Balmer lines. Note that their high $\logg=4.4$, obtained by forcing ionization equilibrium of Fe lines, is reasonably expected for the temperature adopted. They obtain $\afe=6.09$ while for their $\teff$ our \ion{Fe}{i} lines suggest $\afe=5.90$. Axer et~al. noticed that due to systematic differences in their E.W. measurements compared to at least two previous studies, their abundances are probably overestimated by about 0.15 dex. The S/N in this study is significantly lower than in most others. If we correct for this likely error in the E.W. measurements, then $\afe=5.94$, in good agreement with our result. However, the $gf$ values used by these authors were determined using the solar spectrum so the good agreement may be fortuitous.

\item \textit{Spite et~al. (1994)} do not give enough details to make a fair comparison. Their $\teff=5950$~K and $\logg=3.30$ values are from the excitation and ionization balance conditions, but no details are given about the line list and atomic data. No solar $\afe$ is given either.

\item \textit{Th\'evenin \& Idiart (1999)} used the Th\'evenin (1998) catalog as the source for their LTE parameters but no details on their determination are given, except that it is a re-analysis of literature data. They derive non-LTE corrections to the Fe abundances, which amount to about 0.2 dex in the case of BD~+17~4708. According to the authors, only the \ion{Fe}{i} lines suffer from significant deviations from LTE. With their non-LTE corrections, the iron abundance increases from $\afe=5.71$ to 5.92. Our LTE abundance for their $\teff$ is $\afe=5.81$, from both \ion{Fe}{i} and \ion{Fe}{ii} lines (within 0.02 dex). The 0.1 dex difference in the LTE abundance is likely due to the use of solar $gf$ values in their work. As it is discussed in \S\ref{s:iron}, the increase in the \ion{Fe}{i} abundance due to the non-LTE effects predicted by these authors worsens the ionization balance problem we find.

\item \textit{Boesgaard et~al. (1999)} used two $\teff$ scales, those by King (1993, K93) and Carney (1983a,b; C83), which result in 6091~K and 5956~K, respectively. The former is based on the modeling of the H$\alpha$ line while the second is essentially the Peterson (1981) $\teff$ scale. The low $\teff$ obtained in this way may be due to several factors, for example: 1) missing opacity in older Kurucz models which leads to overestimated UV and visible fluxes and thus lower $\teff$ to match the observations, 2) metallicity effects not properly accounted for in the $\teff$ vs. color calibrations, which generally result in low temperatures for metal-poor warm stars if the calibration is constructed mainly with solar metallicity stars, 3) the zero point correction to the $\teff$ scale. Their $\feh$ values have been taken from the literature but put onto the same scale by using the same solar iron abundance. They find $\afe=5.78$ with the K93 scale and $\afe=5.70$ with the C83 scale. These values are both 0.10 dex larger than ours.

\item \textit{Fulbright (2000)} performed a classical spectroscopic abundance analysis, deriving $\teff$ from the excitation equilibrium of \ion{Fe}{i} lines condition and then setting $\logg$ from the ionization balance condition using \ion{Fe}{ii} lines. The process is iterative but it starts with the $\teff$ estimate. Given that $\logg$ has a smaller effect on the $\ion{Fe}{i}$ abundances, the resulting temperatures and metallicities are mostly affected by errors in the \ion{Fe}{i} line modeling. The complete line list used in this study is published and a throughout comparison can be made. In fact, we reproduced this analysis using the same models (Kurucz overshoot) and atomic data. Our resulting $A_\mathrm{Fe}$ vs. E.P. relations are shown in Fig.~\ref{f:ful}. The $gf$ values given by Fulbright have been empirically corrected so it is not surprising to find a smaller scatter compared to our results (we remind the reader that we avoided this type of corrections given that they may artificially reduce the impact of model uncertainties). As shown in Fig.~\ref{f:ful}, with the $\teff$ adopted by Fulbright (6025~K), a small trend with E.P. still remains but it disappears for $\teff=6190$~K. In Fig.~\ref{f:ful} we also show the slope of the $A_\mathrm{Fe}$ vs. E.P. trends ($\epsilon$) as a function of $\teff$ (for comparison purposes we also show our $\epsilon$ values). This $\teff$ increase degrades the almost perfect ionization balance obtained with $\teff=6025$~K (where the \ion{Fe}{i} and \ion{Fe}{ii} abundances agree within 0.03 dex compared to a difference of about 0.08 dex with the high $\teff$). Thus, it seems that the small deviation from excitation balance was sacrificed to almost perfectly satisfy the ionization balance condition. Clearly, the line-to-line scatter in this analysis is still too large as to use the excitation/ionization balance conditions as good $\teff/\logg$ indicators, i.e., there will be always room to alter the $\teff$ and $\logg$ values by changing the criteria for excitation and ionization balance, which are not going to be satisfied simultaneously due to model limitations. Note also that the E.P. coverage of the Fulbright line list is about 2 eV shorter than ours (at least for the lines that can be reasonably well analyzed). The Fulbright analysis suggests $A_\mathrm{Fe}\simeq5.90$, which is larger by about 0.1 dex than our $A_\mathrm{Fe}$, most likely due to the use of Kurucz overshoot instead of no-overshoot models. Despite the empirical corrections, the $gf$ scale of Fulbright is in good agreement with the one used in this work. The use of the Barklem et~al. (2000) damping constants instead of the modified Uns\"old approximation values used by Fulbright does not change the abundances by more than 0.01 dex given that the lines are relatively weak so this does not explain the excitation balance discrepancy. Neither does the use of no-overshoot models.

\begin{figure}
 \centering
 \includegraphics[width=8.0cm]{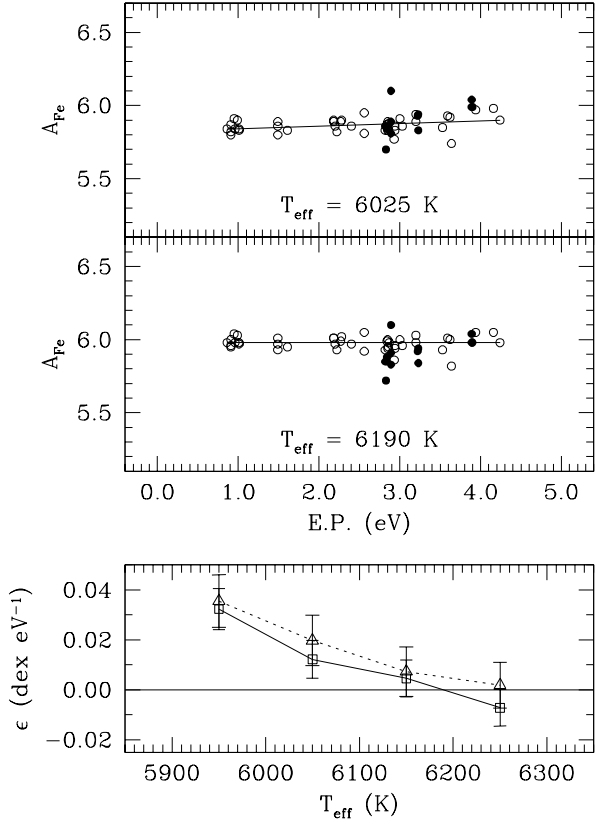}
 \caption{Top panel: abundance of iron as a function of excitation potential for two different effective temperatures as inferred using the Fulbright (2000) line list. Open circles: \ion{Fe}{i} lines, filled circles: \ion{Fe}{ii} lines. The solid lines are linear fits to the \ion{Fe}{i} data only. Bottom panel: slope of the $A_\mathrm{Fe}$ vs. E.P. trend as a function of $\teff$ using the Fulbright (2000) line list (squares and solid line) and our line list (triangles and dotted line). One-sigma ($1\sigma$) errors for the slopes are also shown.}
 \label{f:ful}
\end{figure}

\item \textit{Mishenina et~al. (2000)} obtained $\teff=5943$~K from the Alonso et~al. (1996) IRFM temperature scale and 6000~K from their fitting of models to the H$\alpha$ line. Both values need to be corrected upwards due to reddening and probably observational errors. They derive $\afe=5.90$, in reasonable agreement with our $\afe=5.84$ (after correcting for $\teff$ and solar abundances). Their $\logg=4.0$ value was inferred by forcing ionization equilibrium of Fe lines. However, their mean \ion{Fe}{i} and \ion{Fe}{ii} abundances have uncertainties of 0.11 and 0.17 dex respectively. Although they use the correct $\logg$, the error bar in their derived value should be very large and the claimed agreement between \ion{Fe}{i} and \ion{Fe}{ii} abundances is questionable.

\item \textit{Ryan et~al. (2001)} determined $\teff=5983$~K by using several color calibrations, including that by Magain (1987), in principle all giving essentially the same $\teff$. It is interesting, then, to find that even adopting $E(B-V)=0.01$, they find a low $\teff$ with a $\teff$ scale consistent with that by Magain (1987). This is likely because Ryan et~al. explicitly avoided the high temperatures predicted by the IRFM (see \S5.2 in Ryan et~al. 1999), claiming an unphysical nature of the color-calibrations by Alonso et~al. (1996), which have, however, been recently confirmed by Ram\'{\i}rez \& Mel\'endez (2005b, their \S4.4). The Ryan et~al. $\feh$ and $\logg$ values were taken from the literature.

\item \textit{Simmerer et~al. (2004)} determined $\teff$ from the Alonso et~al. (1996) calibrations but with $E(B-V)=0$, hence the low $\teff\simeq5941$~K. They inferred $\logg=3.98$ both from the \textit{Hipparcos} parallax of the star and the ionization balance condition of iron lines, which is in good agreement with our results for the low $\teff$. Their iron abundance is $\afe=5.92$ while ours for the same parameters is $\afe=5.82$. The 0.1~dex difference between these abundances is probably due to the use of very weak lines ($\mathrm{E.W.}<10$~m\AA\ in most cases), which suffer from errors due to noise and continuum placement. Note that the $gf$ values used in their work are on the same scale as ours (i.e., they were obtained from the same sources), but the lines are not the same.

\item \textit{Nissen et~al. (2004)} obtained $\teff=5943$~K, also from the Alonso et~al. (1996) calibrations with zero reddening (as inferred from the Schuster \& Nissen 1989 calibration) and $\logg=3.97$ from the \textit{Hipparcos} parallax. They find $\afe=5.89$, in reasonable agreement with our $\afe=5.82$. Note that they used MARCS instead of Kurucz models. Most of their iron lines are in the blue ($\lambda<4600$~\AA) so it is not possible to make a line-by-line comparison.

\item \textit{Mel\'endez \& Ram\'{\i}rez (2004)} used literature values for $\feh$ and $\logg$. Their $\teff=6154$~K is from the Ram\'{\i}rez \& Mel\'endez (2005b) IRFM temperature vs. color calibrations, using $E(B-V)\simeq0.02$. Note that for $E(B-V)=0$, the IRFM temperature vs. color calibrations suggest a lower $\teff\simeq6050$~K (Fig.~\ref{f:teffcolor}) while application of the IRFM for this star results in 5955~K (Ram\'{\i}rez \& Mel\'endez 2005a). At $E(B-V)=0.01$, however, the $\teff$ from the color calibrations is in good agreement with that found in our work.

\item \textit{Asplund et~al. (2005)} determined $\teff=6183$~K from fitting of the wings of the H$\alpha$ line using essentially the same atomic data that we used in \S\ref{s:balmer} but MARCS models instead of Kurucz. The effect of using a different model atmosphere is negligible given that we find a very similar H$\alpha$ temperature (\S\ref{s:balmer}). Their $\logg$ value is inferred from the \textit{Hipparcos} parallax of the star. They derive $\feh=-1.51$ ($\afe=5.99$) from both \ion{Fe}{i} and \ion{Fe}{ii} lines, which gave almost the same value for this star (but they find a mean systematic difference of 0.08 dex with the \ion{Fe}{i} lines giving lower abundances in their complete sample). The use of the MARCS model adopted by Asplund et~al. increases the \ion{Fe}{ii} abundance by 0.06 dex with respect to the abundance we derived with the Kurucz model but the \ion{Fe}{i} abundance remains unchanged. Asplund et~al. used only weak \ion{Fe}{i} lines with $gf$ values from O'Brian et~al. (1991) and \ion{Fe}{ii} lines for which $gf$ values are from Bi\'emont et~al. (1994). When we restrict our line list to lines with these characteristics, the \ion{Fe}{i} abundance reduces by 0.03 dex while the \ion{Fe}{ii} abundance increases by 0.02 dex. Thus, the mean (\ion{Fe}{i}--\ion{Fe}{ii}) difference reduces from 0.15 dex when using our line list, adopted atomic data, and a Kurucz model to 0.03 dex if a MARCS model and a line selection similar to that of Asplund et~al. is made.

\end{itemize}

In summary, the effective temperature differences between the one obtained in this paper and those given in the literature can be explained by ignored reddening, limitations of older models, errors in the observations and basic data in the H$\alpha$ line modeling, and the nature of the classical spectroscopic analysis. In most cases we were able to reproduce the literature abundances with corrections due to solar $\afe$ and $\teff$ differences. In the other cases, the remaining differences (0.1 dex or less) can be reasonably explained by the atomic data adopted and/or different model atmospheres. This means that the codes for LTE analyses are free of major bugs. Our $\afe$ values, with conservative error bars, are very robust given that our line list is accurate in an absolute sense


\begin{thebibliography}{}

\bibitem[]{} Allende Prieto, C.
             2001, The Spectrum of the Th-Ar Hollow-Cathode Lamp Used with the 2dcoude Spectrograph,
             McDonald Observatory Technical Note (astro-ph/0111172),
             http://hebe.as.utexas.edu/2dcoude/thar/

\bibitem[]{} Allende Prieto, C., \& Lambert, D. L.
             2000, \aj, 119, 2445

\bibitem[]{} Allende Prieto, C., Asplund, M., Garc\'{\i}a L\'opez, R. J., \& Lambert, D.~L.
             2002, \apj, 567, 544

\bibitem[]{} Allende Prieto, C., Lambert, D. L., Hubeny, I., \& Lanz, T.
             2003a, \apj, 147, 363

\bibitem[]{} Allende Prieto, C., Hubeny, I., \& Lambert, D. L.
             2003b, \apj, 591, 1192

\bibitem[]{} Allende Prieto, C., Barklem, P. S., Lambert, D. L., \& Cunha, K.
             2004a, \aap, 420, 183

\bibitem[]{} Allende Prieto, C., Asplund, M., \& Fabiani Bendicho, P.
             2004b, \aap, 423, 1109

\bibitem[]{} Alonso, A., Arribas, S., \& Mart\'{\i}nez-Roger, C.
             1995, \aap, 297, 197

\bibitem[]{} Alonso, A., Arribas, S., \& Mart\'{\i}nez-Roger, C.
             1996, \aaps, 117, 227

\bibitem[]{} Arenou, F., Grenon, M., \& Gomez, A.
             1992, \aap, 258, 104

\bibitem[]{} Arnone, E., Ryan, S. G., Argast, D., et~al.
             2005, \aap, 430, 507

\bibitem[]{} Asplund, M.
             2005, \araa, 43, 481

\bibitem[]{} Asplund, M., \& Garc\'{\i}a Perez, A. E.
             2001, \aap, 372, 601

\bibitem[]{} Asplund, M., Grevesse, N., \& Sauval, A. J.
             2005, in ASP Conf. Ser. 336, Cosmic Abundances as Records of Stellar Evolution 
             and Nucleosynthesis, ed. T. G. Barnes \& F. Bash (San Francisco, ASP), p. 25

\bibitem[]{} Asplund, M., Lambert, D. L., Nissen, P. E., et~al.
             2006, \apj, 644, 229

\bibitem[]{} Athay, R. G., \& Lites, B. W.
             1972, \apj, 176, 809

\bibitem[]{} Axer, M., Fuhrmann, K., \& Gehren, T.
             1994, \aap, 291, 895

\bibitem[]{} Balachandran, S. C., \& Bell, R. A.
             1998, Nature, 392, 791

\bibitem[]{} Barklem, P. S., \& Aspelund-Johansson, J.
             2005, \aap, 435, 373

\bibitem[]{} Barklem, P. S., Piskunov, N., \& O'Mara, B. J.
             2000, \aaps, 142, 467

\bibitem[]{} Barklem, P. S.,  Stempels, H. C., Allende Prieto, C., et~al.
             2002, \aap, 385, 951

\bibitem[]{} Barklem, P. S.,  Christlieb, N., Beers, T. C., et~al.
             2005, \aap, 439, 129

\bibitem[]{} Bell, R. A., Paltoglou, G., \& Tripicco, M. J.
             1994, \mnras, 268, 771

\bibitem[]{} Bensby, T., Feltzing, S., \& Lundstr\"om, I.
             2003, \aap, 410, 527

\bibitem[]{} Bertelli, G., Bressan, A., Chiosi, C., et~al.
             1994, \aaps, 106, 275

\bibitem[]{} Bi\'emont, E., Baudoux, M., Kurucz, R. L., et~al.
             1991, \aap, 249, 539

\bibitem[]{} Boesgaard, A. M., King, J. R., Deliyannis, C., \& Vogt, S. S.
             1999, \aj, 117, 492

\bibitem[]{} Bohlin, R. C., \& Gilliland, R. L.
             2004a, \aj, 127, 3508

\bibitem[]{} Bohlin, R. C., \& Gilliland, R. L.
             2004b, \aj, 128, 3053 (BL04b)

\bibitem[]{} Bohlin, R. C., Savage, B. D., \& Drake, J. F.
             1978, \apj, 224, 132

\bibitem[]{} Bond, H. E.
             1980, \apjs, 44, 517

\bibitem[]{} Burstein, D., \& Heiles, C.
             1978, \apj, 225, 40

\bibitem[]{} Carney, B. W.
             1983a, \aj, 88, 610  (C83)
             
\bibitem[]{} Carney, B. W.
             1983b, \aj, 88, 623 (C83)

\bibitem[]{} Carretta, E., Gratton, R. G., \& Sneden, C.
             2000, \aap, 356, 238
             
\bibitem[]{} Cayrel de Strobel, G., Soubiran, C., \& Ralite, N.
             2001, \aap, 373, 159

\bibitem[]{} Chen, B., Vergely, J. L., Valette, B., \& Carraro, G.
             1998, \aap, 336, 137

\bibitem[]{} Cowley, C. R.
             1971, The Observatory, 91, 139

\bibitem[]{} Drawin, H. W.
             1968, Z. Phys., 211, 404

\bibitem[]{} Edvardsson, B.
             1988, \aap, 190, 148

\bibitem[]{} Eriksson, K., \& Toft, S. C.
             1979, \aap, 71, 178

\bibitem[]{} Ferlet, R., Vidal-Madjar, A., \& Gry, C.
             1985, \apj, 298, 838

\bibitem[]{} Fitzgerald, M. P.
             1968, \aj, 73, 983

\bibitem[]{} Fitzpatrick, E. L.
             1999, \pasp, 111, 63

\bibitem[]{} Fuhrmann, K., Axer, M., \& Gehren, T.
             1995, \aap, 301, 492

\bibitem[]{} Fulbright, J. P.
             2000, \aj, 120, 1841

\bibitem[]{} Fulbright, J. P., \& Johnson, J. A.
             2003, \apj, 595, 1154

\bibitem[]{} Fukugita, M., Ichikawa, T., Gunn, J. E., et~al.
             1996, \aj, 111, 1748

\bibitem[]{} Garc\'{\i}a-Gil, A., Garc\'{\i}a-Lopez, R. J., Allende Prieto, C., \& Hubeny, I.
             2005, \apj, 623, 460

\bibitem[]{} Gratton, R. G., Carretta, E., Matteucci, F., \& Sneden, C.
             2000, \aap, 358, 671

\bibitem[]{} Gray, D.
             1992, The observation and analysis of stellar photospheres, 2nd ed.
             (Cambridge University Press) p. 207

\bibitem[]{} Grevesse, N., \& Sauval, A. J.
             1998, Space Science Reviews, 85, 161

\bibitem[]{} Gunn, J. E., Carr, M., Rockosi, C., et al.
             1998, \aj, 116, 3040

\bibitem[]{} Gunn, J. E., Siegmund, W. A., Mannery, E. J., et al.
             2006, \aj, 131, 2332

\bibitem[]{} Hakkila, J., Myers, J. M., Stidham, B. J., Hartmann, D. H.
             1997, \aj, 114, 2043

\bibitem[]{} Hubeny, I.
             1988, Computer Phys. Comm., 52, 103

\bibitem[]{} Hubeny, I., \& Lanz, T.
             1995, \apj, 439, 875

\bibitem[]{} Idiart, T., \& Th\'evenin, F.
             2000, \apj, 541, 207

\bibitem[]{} J\o rgensen, I.
             1994, \pasp, 106, 967

\bibitem[]{} Kim, Y. C., Demarque, P., Yi, S. K., \& Alexander, D. R.
             2002, \apjs, 143, 499

\bibitem[]{} King, J. R.
             1993, \aj, 106, 1206 (K93)

\bibitem[]{} Kiselman, D.
             1993, \aap, 275, 269

\bibitem[]{} Korn, A. J., Shi, J., \& Gehren, T.
             2003, \aap, 407, 691

\bibitem[]{} Kurucz, R. L.
             1970, SAO Special Report No. 308

\bibitem[]{} Kurucz, R. L.
             1979, \apjs, 40, 1

\bibitem[]{} Kurucz, R. L., Furenlid, I., Brault, J., \& Testerman, L.
             1984, NSO Atlas No.~1: Solar Flux Atlas from 296 to 1300 nm, Sunspot, NSO

\bibitem[]{} Lallement, R., Bertin, P., Chassefi\`{e}re, \& Scott, N.
             1993, \aap, 271, 734

\bibitem[]{} Lallement, R., Welsh, B. Y., Vergely, J. L., et~al.
             2003, \aap, 441, 447

\bibitem[]{} Lambert, D. L., Heath, J. E., Lemke, M., \& Drake, J.
             1996, \apjs, 103, 183

\bibitem[]{} Latham, D. W., Mazeh, T., Carney, B. W., et~al.
             1988, \aj, 96, 567

\bibitem[]{} Li, J., \& Zhao, G.
             2004, ChJAA, 4, 75

\bibitem[]{} Magain, P.
             1987, \aap, 181, 323

\bibitem[]{} Magain, P.
             1989, \aap, 209, 211

\bibitem[]{} Mel\'endez, J., \& Ram\'{\i}rez, I.
             2004, \apj, 615, L33

\bibitem[]{} Mel\'endez, J., Shchukina, N. G., Vasiljeva, I. E., \& Ram\'{\i}rez, I.
             2006, \apj, in press (astro-ph/0601256)

\bibitem[]{} Mishenina, T. V., Korotin, S. A., Klochkova, V. G., \& Panchuk, V. E.
             2000, \aap, 353, 978

\bibitem[]{} Morton, D. C.
             2003, \apjs, 149, 205

\bibitem[]{} Nissen, P. E., Primas, F., Asplund, M., \& Lambert, D. L.
             2002, \aap, 390, 235
             
\bibitem[]{} Nissen, P. E., Chen, Y. Q., Asplund, M., \& Pettini, M.
             2004, \aap, 415, 993

\bibitem[]{} Nordstr\"om, B., Mayor, M., Andersen, J., et~al.
             2004, \aap, 418, 989

\bibitem[]{} O'Brian, T. R., Wickliffe, M. E., Lawler, J. E., et~al.
             1991, J. Opt. Soc. Am., B8, 1185

\bibitem[]{} Oke, J. B., \& Gunn, J. E.
             1983, \apj, 266, 713

\bibitem[]{} Peterson, R. C.
             1981, \apj, 244, 989

\bibitem[]{} Pont, F., \& Eyer, L.
             2004, \mnras, 351, 487

\bibitem[]{} Rachford, B. L., Snow, T. P., Tumlinson, J., et~al.
             2002, \apj, 577, 221

\bibitem[]{} Ram\'{\i}rez, I., \& Mel\'endez, J.
             2005a, \apj, 626, 446

\bibitem[]{} Ram\'{\i}rez, I., \& Mel\'endez, J.
             2005b, \apj, 626, 465

\bibitem[]{} Ram\'{\i}rez, I., Allende Prieto, C., \& Lambert, D. L.
             2006, in preparation

\bibitem[]{} Rebolo, R., Beckman, J. E., \& Molaro, P.
             1988, \aap, 192, 192

\bibitem[]{} Redfield, S., \& Linsky, J. L.
             2004a, \apj, 602, 776

\bibitem[]{} Redfield, S., \& Linsky, J. L.
             2004b, \apj, 613, 1004

\bibitem[]{} Reetz, J.
             1999, Ph.D. Thesis, Ludwig-Maximilians Univ.

\bibitem[]{} Rufener, F., \& Nicolet, B.
             1988, \aap, 206, 357

\bibitem[]{} Ryan, S. G., Norris, J. E., \& Beers, T. C.
             1999, \apj, 523, 654

\bibitem[]{} Ryan, S. G., Kajino, T., Beers, T. C., et~al.
             2001, \apj, 549, 55

\bibitem[]{} Salaris, M., Chieffi, A., \& Straniero, O.
             1993, \apj, 414, 580

\bibitem[]{} Schlegel, D. J., Finkbeiner, D. P., \& Davis, M.
             1998, \apj, 500, 525

\bibitem[]{} Schuster, W. J., \& Nissen, P. E. 
             1989, \aap, 221, 65

\bibitem[]{} Shchukina, N. G., Trujillo Bueno, J., \& Asplund, M.
             2005, \apj, 618, 939

\bibitem[]{} Simmerer, J., Sneden, C., Cowan, J. J., et~al.
             2004, \apj, 617, 1091

\bibitem[]{} Smith, J. A., Tucker, D. L., Kent, S., et~al.
             2002, \aj, 123, 2121

\bibitem[]{} Sneden, C.
             1973, Ph.D. thesis, Univ. Texas at Austin

\bibitem[]{} Spite, M., Pasquini, L., \& Spite, F.
             1994, \aap, 290, 217

\bibitem[]{} Steenbock, W., \& Holweger, H.
             1984, \aap, 130, 319

\bibitem[]{} Stehl\'e, C., \& Hutcheon, R.
             1999, \aaps, 140, 93

\bibitem[]{} Strauss, M., \& Gunn, J. E.
             2001, Technical Note available from http://www.sdss.org/dr3/instruments/imager

\bibitem[]{} Th\'evenin, F.
             1998, Chemical Abundances in Late-Type Stars.
             VizieR On-line Data Catalog: III/193

\bibitem[]{} Th\'evenin, F., \& Idiart, T. P.
             1999, \apj, 521, 753

\bibitem[]{} Tull, R. G., MacQueen, P. J., Sneden, C., \& Lambert, D. L.
             1995, \pasp, 107, 251

\bibitem[]{} Vandenberg, D. A., Stetson, P. B.
             1991, \aj, 102, 1043
 
\bibitem[]{} VandenBerg, D. A., Swenson, F. J., Rogers, F. J., et~al.
             2000, \apj, 532, 430

\bibitem[]{} Welty, D. E., Hobbs, L. M., \& Kulkarni, V. P.
             1994, \apj, 436, 152

\bibitem[]{} Wheeler, J. C., Sneden, C., \& Truran, J. W.
             1989, \araa, 27, 279

\bibitem[]{} Zhou, X., Jiang, Z., Xue, S., et~al.
             2001, ChJAA, 1, 372

\end{thebibliography}
\end{document}